\documentclass[twocolumn,superscriptaddress]{revtex4-1}  
\usepackage{graphicx}  
\usepackage{dcolumn}   
\usepackage{multirow}
\usepackage{color,soul}
\usepackage{amssymb}   
\usepackage{array}
\usepackage{amsmath}
\usepackage{tabularx}
\usepackage{natbib}
\usepackage{notes2bib}
\usepackage{epstopdf}
\usepackage{mathrsfs} 
\usepackage[normalem]{ulem} 
\usepackage[hidelinks]{hyperref}
\usepackage{booktabs}

\usepackage[table,xcdraw]{xcolor}
\definecolor{light-gray}{gray}{0.9}
\usepackage{soulpos}
\newcommand{\hdcell}[1]{\cellcolor{light-gray}{\bf #1}}

\usepackage{xspace}

\newcommand{\figref}[1]{Figure~\ref{fig:#1}\xspace}
\newcommand{\tabref}[1]{Table~\ref{tab:#1}\xspace}
\newcommand{\secref}[1]{Section~\ref{sec:#1}\xspace}
\newcommand{\secpageref}[1]{Page~\pageref{sec:#1}\xspace}

\DeclareMathAlphabet{\mathtensor}{OT1}{cmss}{bx}{n}

\let\savedCaption=\caption
\renewcommand*{\caption}[2][\shortcaption]{%
  \def\shortcaption{\footnotesize\bf\sf #2}
  \savedCaption[#1]{\footnotesize\bf\sf #2}}

\begin{document}

\title{Machine Learning Approach to RF Transmitter Identification}

\author{K. Youssef}
\author{L.-S. Bouchard}
\affiliation{Department of Chemistry and Biochemistry, Bioengineering and California NanoSystems Institute, University of California, Los Angeles, California 90095, USA}
\author{K.Z. Haigh}
\author{H. Krovi}
\author{J. Silovsky}
\author{C.P. Vander Valk}
\affiliation{Raytheon BBN Technologies Corp., 10 Moulton Street, Cambridge, MA 02138, USA}

\date{\today}

\begin{abstract}



  \fbox{
    \begin{minipage}{5in}
      This document does not contain technology or technical data controlled under either the U.S. International Traffic in Arms Regulations or the U.S. Export Administration Regulations.
      \end{minipage}
  }

With the development and widespread use of wireless devices in recent years (mobile phones, Internet of Things, Wi-Fi), the electromagnetic spectrum has become extremely crowded.  In order to counter security threats posed by rogue or unknown transmitters, it is important to identify RF transmitters not by the data content of the transmissions but based on the intrinsic physical characteristics of the transmitters.  RF waveforms represent a particular challenge because of the extremely high data rates involved and the potentially large number of transmitters present in a given location.  These factors outline the need for rapid fingerprinting and identification methods that go beyond the traditional hand-engineered approaches.  In this study, we investigate the use of machine learning (ML) strategies to the classification and identification problems, and the use of wavelets to reduce the amount of data required.  Four different ML strategies are evaluated: deep neural nets (DNN), convolutional neural nets (CNN), support vector machines (SVM), and multi-stage training (MST) using accelerated Levenberg-Marquardt (A-LM) updates. The A-LM MST method preconditioned by wavelets was by far the most accurate, achieving 100\% classification accuracy of transmitters, as tested using data originating from 12 different transmitters.  We discuss strategies for extension of MST to a much larger number of transmitters. 
 \end{abstract}

\maketitle

\section{Introduction}

Due to the very large number of electronic devices in today's environment, the RF spectrum is very crowded.  Examples of devices commonly encountered that emit radio waves include cordless phones, cell phones, microwave ovens, wireless audio and video transmitters, motion detectors, WLAN, and cars.  With the advent of the Internet of Things, an even larger swath of devices contributes to the RF emissions, in the form of physical devices, vehicles, and other items embedded with electronics, software, sensors, actuators, and network connectivity, which enable these objects to communicate by transmitting and receiving data.  A large number of communication protocols currently operate in different RF bands (ANT+, Bluetooth, cellular LTE, IEEE 802.15.4 and 802.22, ISA100a, ISM, NFC, 6LoWPAN, UWB, IEEE's Wi-Fi 802.11, Wireless HART, WirelessHD, WirelessUSB, ZigBee, Z-Wave). Many of these devices are wildly insecure for a variety of reasons~\cite{FLU01,KEL15,GOO15}. It is thus imperative to solve the security vulnerabilities or identify and counter the attacks.  Vulnerabilities in the broader sense not only include attacks to the devices, but also false impersonations of these devices, for example, by rogue transmitters.  The rapid identification of threats from unknown signals is of paramount importance.

\begin{figure}
  \fbox{
  \begin{minipage}{0.95\columnwidth}
    \centerline{\bf Reader's Guide}
    \raggedright
    \bgroup
    \def\arraystretch{1.5}
    \begin{tabular}{ p{0.5\columnwidth} p{0.45\columnwidth}}
      Which ML models accurately distinguish known transmitters?  &      {\bf Classification:}\newline\secref{classification},  page~\pageref{sec:classification}\\
      Does MST owe its superior performance to the multi-stage training
        strategy, or to our use of second-order (LM) training updates? & {\bf First- vs Second-order Training:}\newline\secref{first-vs-second}, page~\pageref{sec:first-vs-second} \\
      How well can the models extend to capture new devices?  & {\bf Incremental Learning:}\newline \secref{incremental},  page~\pageref{sec:incremental}\\
      Can we obtain similar performance while keeping network sizes small? &
      {\bf Wavelet Preconditioning:}\newline\secref{wprec}, page~\pageref{sec:wprec} \\
      Can we extend similar performance while keeping network sizes small? &
      {\bf Incremental Learning with Wavelets:}\newline\secref{wprec}, page~\pageref{sec:wprec} \\
    \end{tabular}
    \egroup
  \end{minipage}
  }
\end{figure}

Another important motivation for the development of transmitter identification schemes is the mitigation of problems associated with RF interference.  Because the overlap between the different bands is strong and the number of transmitters can be large, the SNR is often reduced due to RF interference.  RF interference can be generated by almost any device that emits an electro-magnetic signal, from cordless phones to Bluetooth headsets, microwave ovens and even smart meters. The single biggest source of Wi-Fi interference is the local Wi-Fi network because Wi-Fi is a shared medium that operates in the unlicensed ISM bands within the 2.4 GHz to 5 GHz range.  Compounding the problem is the fact that transmitters tend to operate independently.  Thus, the lack of timing alignment results in significant inter-channel interference.  When interference occurs, RF packets are lost and must be retransmitted.  Conventional techniques such as switching frequency bands are insufficient at solving the interference problem.   Methods that more accurately identify the transmitting sources could lead to better schemes for signal separation in a crowded environment.

For identification of threats to aircrafts, Radar Warning Receivers (RWR) typically analyze RF frequency, pulse width, pulse-repetition frequency, modulation (chirp or binary code) on pulses, CW modulations and antenna scan characteristics. In modern RWR systems, the computer determines the closest fit of these parameters to those found in a threat identification table to force an identification.  Even modern RWR systems do not use identification algorithms that go well beyond schemes based on matching these parameters.  Thus, it makes sense to explore more sophisticated techniques. 

In recent years, the field of artificial intelligence (AI) has rapidly grown in popularity due to the development of modern ML techniques, such as deep learning.    Applications of AI to image and speech recognition in everyday life situations are now increasingly common.  In contrast, AI is scarcely used in the RF domain and little work has been done to explore the connection between RF signal processing and ML.   In this study, we extend ML approaches to the RF spectrum domain in order to develop practical applications in emerging spectrum problems, which demand vastly improved discrimination performance over today's hand-engineered RF systems. 

In RF applications, particular challenges exist having to do with the numerous transmission protocols  and the large amounts of data due to the large bandwidths and high data rates involved. This calls for the development of new algorithms capable of addressing these challenges.  While many of the modern ML algorithms were born from the desire to mimic biological systems, RF systems have no biological analogues, and tailored AI strategies are needed.  

\section{Problem Description}

The main task that must be implemented is RF feature learning.   A naive application of AI to the RF spectrum depends on hand-engineered features, which an expert has selected based on the belief that they best describe RF signals pertinent to a specific RF task.   On the other hand, the application of deep learning to other domains has achieved excellent performance in vision~\cite{bib:DL_image} and speech~\cite{bib:DL_speech} problems by learning features similar to those learned by the brain from sensory data. Recently, it has been shown~\cite{DEP17,KAR17,DYS17} that ML of RF features has the potential to transform spectrum problems in a way similar to other domains. Specifically, AI schemes should be capable of learning the appropriate features to describe RF signals and associated properties from training data. Ultimately, ML innovations will result in a new generation of RF systems that can learn from data.
  The development of transmitter identification schemes would help counter security threats and mitigate problems associated with RF interference.

Herein we explored several ML strategies for RF fingerprinting as applied
to the classification and identification of RF Orthogonal Frequency-Division
Multiplexing (OFDM) packets~\cite{ofdm17}:
\begin{itemize}
\itemsep=0pt\parsep=0pt\parskip=0pt

  \item Support Vector Machines (SVM), with two different kernels,
  \item Deep Neural Nets (DNN),
  \item Convolutional Neural Nets (CNN),
    and
  \item Accelerated second-order Levenberg-Marquardt (A-LM)~\cite{YOU17}
      Multi-Stage Training (MST)~\cite{YOU15,YOU15b} (A-LM MST).  A comparison with first-order training is included. 
\end{itemize}
We find that the highest accuracy across a broad range of conditions, including exposure to a more limited training dataset, is achieved using our A-LM MST method.  We validate our methods on experimental data from 12 different OFDM transmitters. Strategies for extensions to much larger numbers of transmitters are discussed, including a promising approach based on the preconditioning of RF packets by decomposition of the RF signal content into wavelets ahead of the ML phase.

\section{Data Preparation Method}

Our sample RF data was collected from six different radios with two transmitters each, for a total of $N_t=12$ transmitters. The radios share power supply and reference oscillator, while the rest of the RF chain differs for each transmitter (digital-to-analog converter, phase-locked loop, mixer, amplifiers and filters).  The RF data, which was stored using the name convention $\langle$radio name$\rangle$\_$\langle$transceiver number$\rangle$\_$\langle$packet number$\rangle$, was captured at 5 MSPS with an Ettus USRP N210 with WBX daughter card. 

We collected 12,000 packets total, 1,000 packets per transmitter.  The packets were generated with pseudo-random values injected through a debug interface to the modem; no MAC address or other identification was included in signal.  The same set of 1,000 packets were transmitted by each radio.
We used a proprietary OFDM protocol, with baseband sample rate of the transmitter of 1.92 MSPS (1.2 MHz bandwidth), 3.75 kHz subcarrier spacing, cyclic prefix length 20 samples, 302 subcarriers with QPSK modulation. Each captured packet was represented by $n=10,000$ time-domain complex-valued I/Q data points.  To reduce ambiguity in describing the data preparation and handling, we denote a time-domain data collection by the complex-valued vector 
$$\vec{f}=(f_1,f_2,\dots,f_n),$$
where $n=10,000$ is the number of time-domain points and $f_i\in \mathbb{C}$, $i=1,\dots,n$.

For each signal packet, we detect the onset by thresholding the real value, $\Re(f_i)$, thereby skipping the first $N_o$ data points where $|\Re(f_i)|<\tau$ for some threshold value~\footnote{We used a threshold $\tau$ of 0.05, although the exact value here is unimportant because this scale is relative. } $\tau>0$ chosen well above the noise, $i<N_o$, and take the next $N$ data points in the waveform, to yield a signal vector $\vec{g}$,
$$\vec{g}=(f_{N_o},f_{N_o+1},\dots,f_{N_o+N-1}).$$ 
This method is referred to as w$N$, where $N$ is varied (e.g., w32, w64, w128, w256, w512, w1024).
For DNN, SVM and MST processing, a vector $\vec{v}$ was constructed by concatenating the real and imaginary parts of $\vec{g}$ into a vector of length $2N$:
$$ \vec{v}=(\Re g_1, \dots,\Re g_N,\Im g_1,\dots,\Im g_N).$$
For CNN processing, a real and imaginary parts were treated as elements of a two-dimensional vector and the input to the network formed as a sequence of $N$ of these vectors. Handling of the signal in case of  wavelet preconditioning will be described in Section~\ref{sec:wprec}.

We explored the effects of training the different ML techniques using different amounts of training vs testing data: 1) 90\%  of the data used for training and 10\%  for testing, for all values of $N$.   This experiment will be denoted as 90/10.  2) 10\% of the data was used for training and 90\% for testing, for all values of $N$. This will be denoted as 10/90. For our dataset of 12 transmitters, each with 1,000 packets captured, 90\% of data corresponds to 10,800 packets and 10\% of data to the remaining 1,200 packets.

\section{Algorithms}

In order to demonstrate the ability of ML to learn features from RF signals, create models that can identify and distinguish different known transmitters, and recognize unknown transmitters to a high degree of accuracy, four different algorithms are investigated: SVM, CNN, DNN and MST. SVM and MST have two configurations each, for a total of six different analyses. These methods and their implementations are described below.

\subsection{Support Vector Machines \label{sec:svm}}

We used the SVM implementation found in Weka~\cite{hall09}. We tested with both  the (a) PolyKernel and (b) the Pearson VII Universal Kernel~\cite{ustun06}. PuK is known to be more effective than PolyKernel, but the Weka implementation is extremely slow. (Our prior work~\cite{haigh2015-embedded,haigh2015-commex} re-implemented PuK so that it would operate efficiently on an embedded platform for Support Vector Regression.) We used Platt's Minimization Optimization (SMO) algorithm~\cite{platt98} to compute the maximum-margin hyperplanes.

\subsection{Deep Neural Nets \label{sec:dnn}}

To set a baseline for neural net models, we used a simple DNN with two fully-connected hidden layers, each with 128 nodes. We used rectified linear units (ReLU) as non-linearity in the hidden layers and sigmoid transfer function in the output layer. Mini-batch size of 32 and Adam optimizer were used in the training.

\subsection{Convolutional Neural Nets \label{sec:cnn}}

Our CNN model is composed of two convolutional layers and one fully connected hidden layer. The first convolutional layer had 64 filters of size 8$\times$2, followed by max-pooling layer with 2$\times$2 pool size. The second convolutional layer had 32 filters of size 16$\times$1, followed by max-pooling layer with 2$\times$1 pool size. The fully connected layer had 128 nodes and ReLU non-linearity. As in the DNN case, we used a sigmoid transfer function for the output layer.

\subsection{Multi-Stage Training \label{sec:mst}}

The MST method for ANN, which was first developed for handling large datasets with limited computing resources in image noise identification and removal~\cite{YOU15,YOU15b}, is applied to the RF identification problem for the first time.  It is an alternative method to achieve deep learning with fewer resources.  We present the MST approach with second-order training in~\secref{second-order-mst} and then compare it to the case of MST with first-order training in~\secref{first-order-mst}.  We begin by reviewing the operational principle of MST because it is not as widespread as other ML methods.

\subsubsection{Training neural networks by multiple stages}

\begin{figure}
\includegraphics[width=0.8\columnwidth]{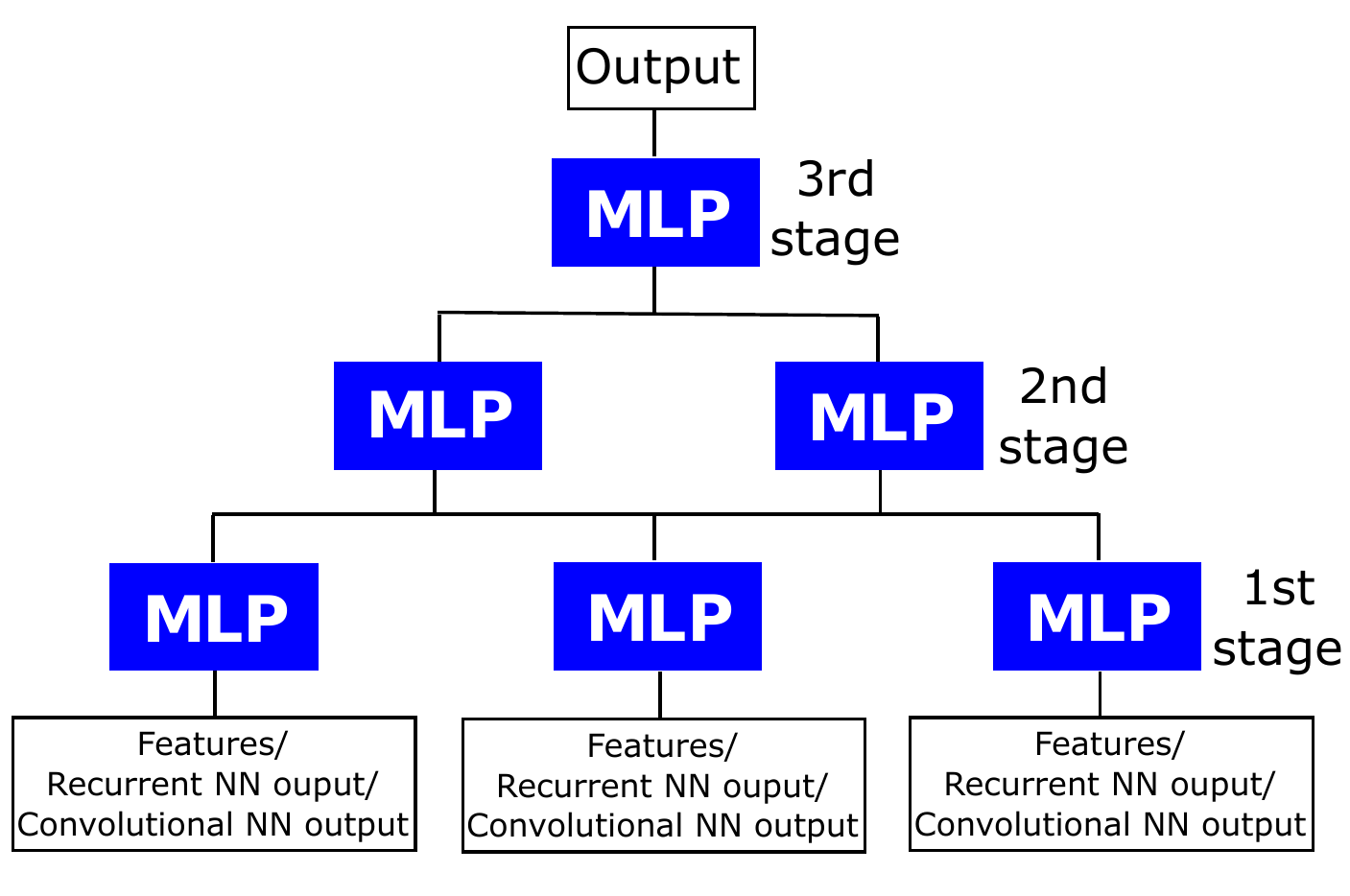}
\caption{\label{fig:MST} The MST method employs groups of MLP organized in a hierarchical fashion.  Outputs of MLPs in the first stage are fed as inputs to MLPs in the second stage.  Outputs of MLPs in the second stage are fed as inputs to MLPs in the third stage, and so on.  While not implemented in this work, in general, the outputs of a stage can be fed into inputs of any higher stage (e.g., outputs of stage 4 could be fed to stage 9 in addition to stage 5).  A front-end can be added to process the input prior to reaching the first stage.  }
\end{figure}

In MST, training is performed in multiple stages, where each stage consists of one or more multi-layer perceptrons (MLP), as shown in Figure~\ref{fig:MST}.  The hierarchical strategy drastically increases the efficiency of training~\cite{YOU15,YOU15b}. The burden of reaching a more global solution of a complex model that can perform well on all variations of input data is divided into multiple simpler models such that each simple model performs well only on a part of the variations in input data. Using subsequent stages, the area of specialization of each model is gradually broadened. The process of reaching a more global minimum becomes much easier after the first stage, since models in the following stages search for combinations of partial solutions of the problem rather than directly finding a complete solution using the raw data.

\begin{figure}
     \includegraphics[width=0.95\columnwidth]{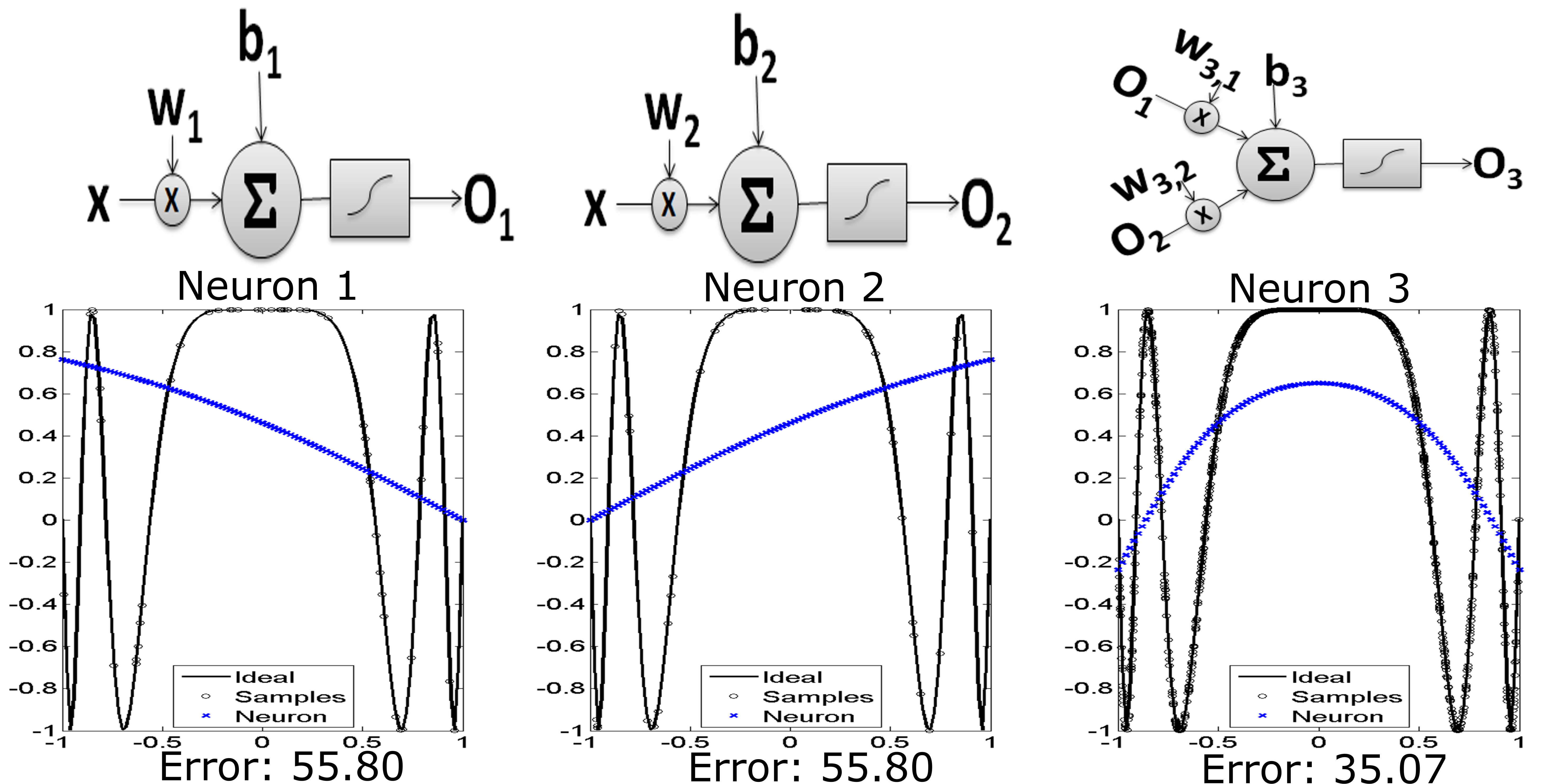}
  \caption{\label{fig:MSTint1}  Toy model illustration of MST training.  In its simplest incarnation, one perceptron (neuron) is trained at a time. Each of
 neurons 1 and 2 are trained to model the ideal response using a set of training samples consisting of the $x$ value as input ($x$-axis) and the corresponding $y$ value as target ($y$-axis), for different input ranges. The number of training samples is increased for Neuron 3, where its inputs are the partial solutions from Neuron 1 and 2 instead of the $x$ value. Neuron 3 finds a combination that keeps the best parts of the partial solutions from Neurons 1 and 2.  {\em Black curve: target function; blue curve: fit result.}
  }
\end{figure}

\begin{figure}
    \includegraphics[width=0.95\columnwidth]{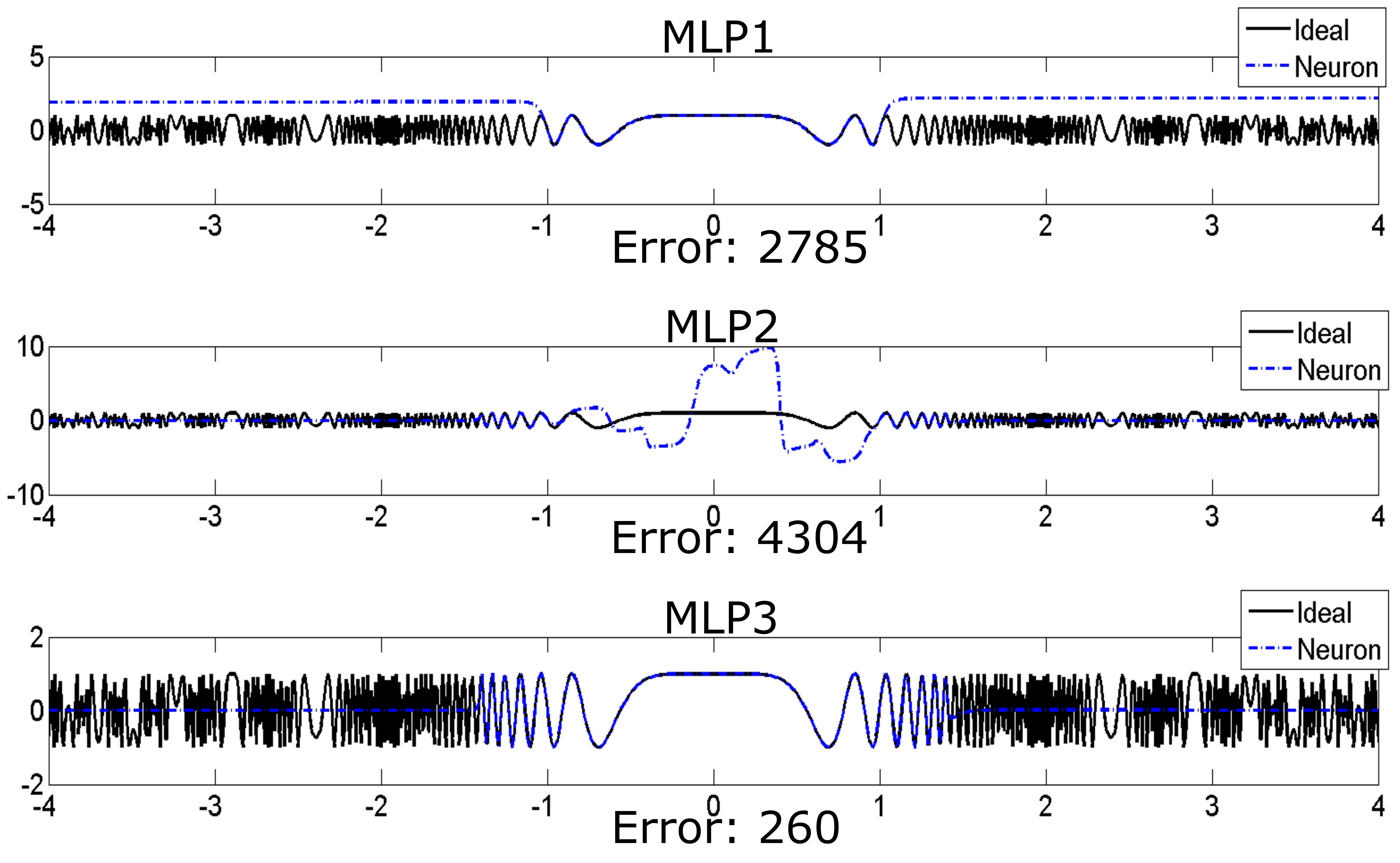} 
  \caption{\label{fig:MSTint2} For more complex functions, single perceptrons are replaced with a network of perceptrons (MLP). Each of MLP 1 and 2 are    trained to model the ideal response using a set of training samples    consisting of the $x$ value as input ($x$-axis) and the corresponding $y$    value as target ($y$-axis), for different input ranges. The number of    training samples and range is increased for MLP 3, where its inputs are the    partial solutions from MLP 1 and 2 instead of the $x$ value. MLP 3 finds a  combination that keeps the best parts of the partial solutions from MLP 1 and 2.    Black curve: target function; blue curve: fit result.
 }
\end{figure}

The level of success of the MST strategy depends largely on assigning the right distribution of training tasks to minimize redundancy within models and increase the diversity of areas of specialization of different models.  When training is done properly, MST can be very efficient, as illustrated in Figures~\ref{fig:MSTint1} and~\ref{fig:MSTint2} for a toy model.  The main idea is to divide training over several smaller MLPs.  This architecture, which is more computationally tractable than training one large MLP, drastically simplifies the approach to deep learning.

We use simple MLP models in the first stage, each trained on a batch consisting of a small part of the training dataset. For example, a training dataset consisting of $N$ training samples can be divided into 20 batches with $N$/10 samples each, noting that batches can have common training samples. For an MST with $M$ MLPs in the first stage, the MLPs are divided into groups of $M$/10 MLPs, where each group is trained using one of the batches. The batch size is progressively increased at higher stages, while the input dimension to each stage is typically decreased. For example, the configuration used herein has stage 1 MLPs with an input size of up to 2,048.  Stage 2 MLPs have an input size of 60, which is the number of MLPs in the first stage. Additionally, by systematically assigning specific stopping criteria to each stage, we gain a level of control over how fast the overall system fits the data, resulting in better overall performance. For example, an MST can be designed with a few stages where a small target error is chosen at the first stage and drastically decreased at successive stages.  Alternatively, a larger target error can be chosen and slowly decreased over more stages, depending on the complexity of the problem and the convergence properties of the training algorithm.  We have shown that MST uses second order methods' ability to yield optimal stopping criteria to produce ANNs with better generalizing ability~\cite{YOU17,YOU15,YOU15b}. These advantages are leveraged here for RF signal classification and identification.

\subsubsection{Second-order updates \label{sec:second-order-mst}}

Feed-forward neural networks such as MLP are typically trained by back-propagation, whereby the weights are updated iteratively using first- or second-order update rules.   First-order updates are generated using the gradient descent method or a variant of it.  Second-order methods are based on the Newton, Gauss-Newton or Levenberg-Marquardt (LM) update rules~\cite{bib:yu,bib:hagen94,bib:hertz91,bib:battiti92}.  LM training yields better results, faster, than first-order methods~\cite{bib:yu,bib:huang06,bib:saravanan}. However, LM cannot be used to train large-scale ANNs, even on modern computers, because of complexity issues~\cite{bib:battiti92}.  Inversion of the Hessian matrix requires $O(N^{2.373})-O(N^3)$ operations, depending on the algorithm used, where $N$ is the network size (i.e. number of parameters).  This is the main computational challenge for LM.  To overcome this problem, we used a variant of the LM update rule termed ``Accelerated LM'' (A-LM), which overcomes the computational challenges associated with LM, and enables us to handle much larger networks and converge faster~\cite{YOU17}.  Apart from computational complexity differences, the end solution-quality result between LM and A-LM, however, is very close.    

\begin{figure*}
    \setlength\tabcolsep{0.5cm}
  \begin{tabular}{ccc}
    \includegraphics[width=0.28\textwidth]{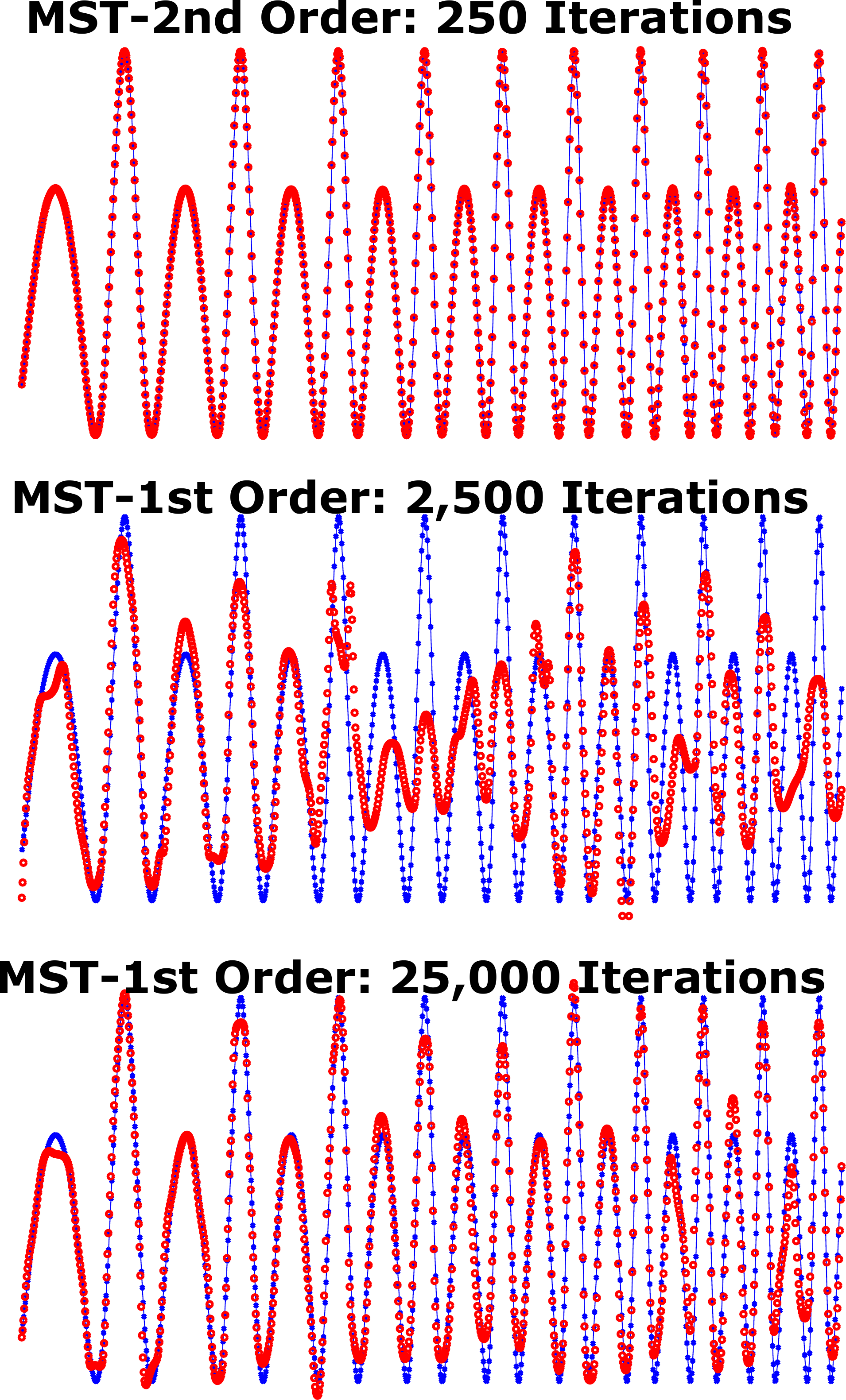} &
    \includegraphics[width=0.28\textwidth]{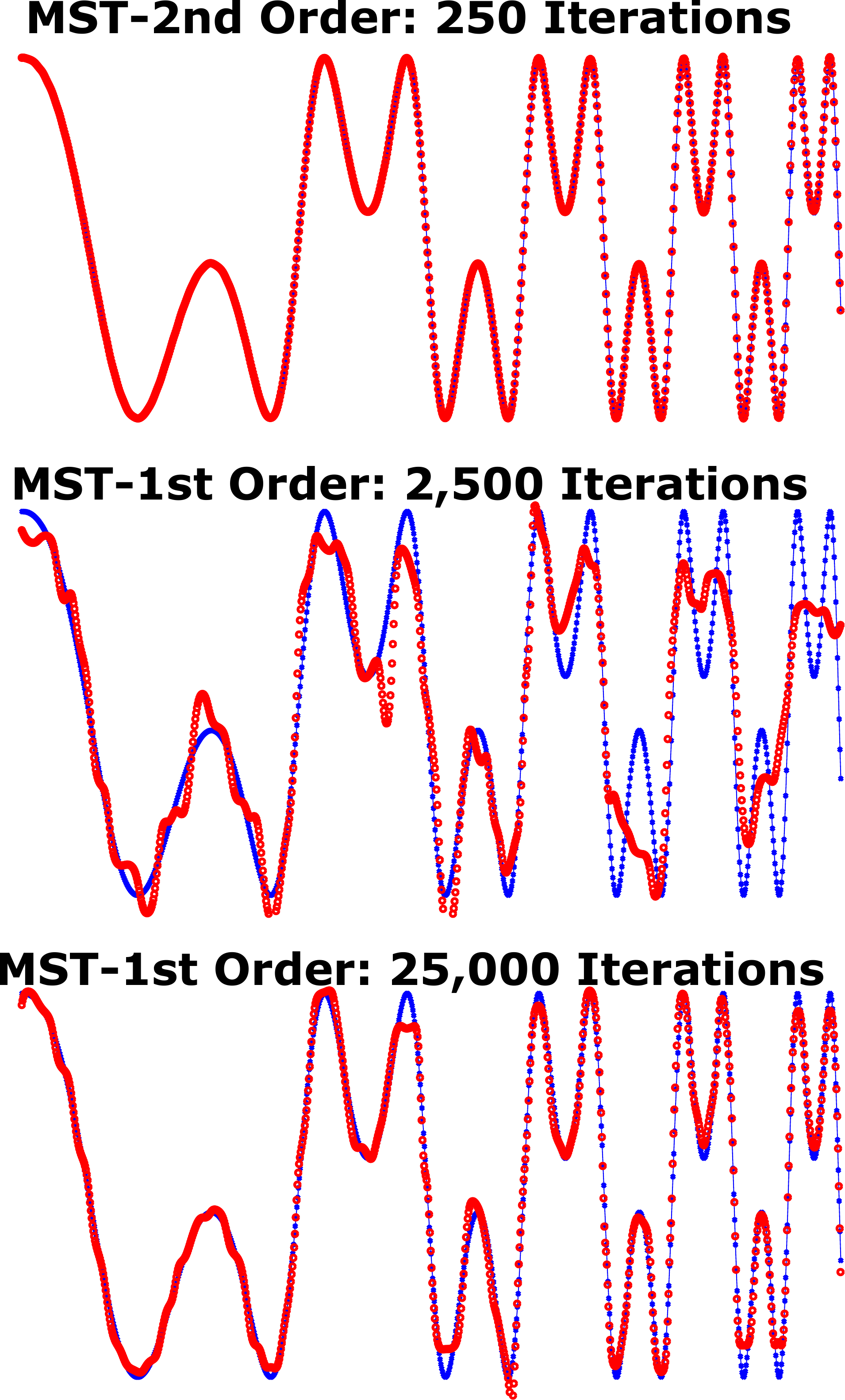} &
    \includegraphics[width=0.28\textwidth]{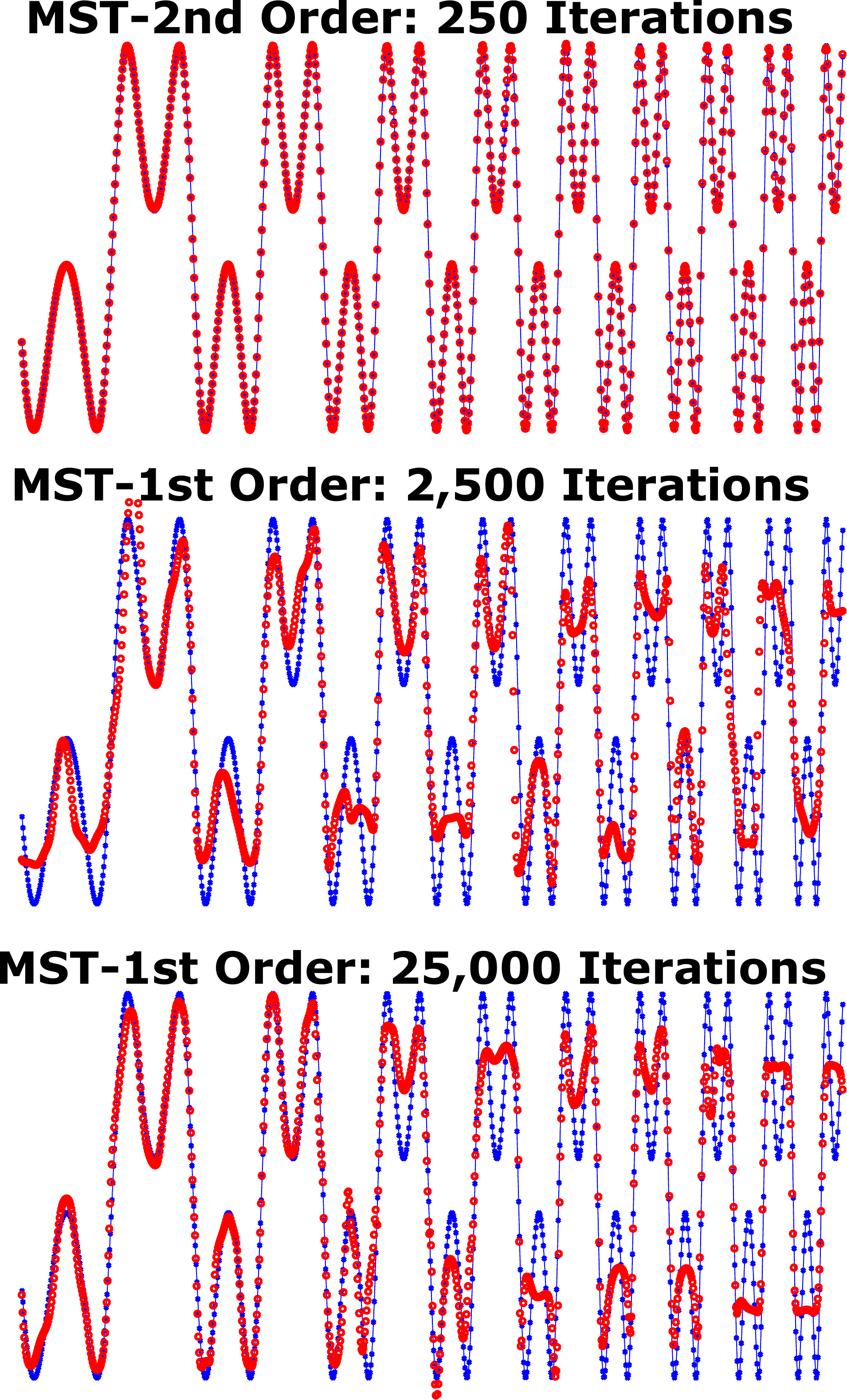} \\
    (a) $f(x)=\cos(5\sin^2(2x))$, $x\in [1,3]$ & (b) $f(x)=\sin(3\cos(x^3))$, $x\in [1,3]$ & (c) $f(x)=\sin(3\cos(x^3))$, $x\in[2,4]$ 
\end{tabular}
\caption{\label{fig:LM_vs_SD} Performance of single MST under first- and second-order training for a curve-fitting task, for the same network architecture and training data set. (a-c) Four different functions are fitted.  The target function is shown in blue whereas the result of the fit is shown in red.  {\em Top row:} second-order A-LM updates (after 250 iterations). {\em Middle and Bottom row:} first-order Steepest-Descent (SD,  with fixed learning rate = 0.01) updates (2,500 and 25,000 iterations, respectively).    For all (a-c) functions, second-order method fully converges after only 250 iterations.  SD converges very slowly, as seen in the bottom row where 25,000 iterations is still not enough to yield a good result.  MST architecture for all experiments: four MLPs in the first stage, one MLP in the second stage, two hidden layers per MLP, 20 neurons per layer. For each function (a-c), the training data set consisted of 5,000 randomly generated values of $x$ from the given range as inputs, and the corresponding values of $y=f(x)$ as targets. The testing data set consisted of 1,000 equally spaced values of $x$ from within the given range, where the MST is required to estimate the corresponding $y$ value. The number of iterations stated is the total number of iterations for all MLPs. Each MLP within the MST was trained for an equal number of iterations for each case, which is equal to the total number of iterations divided by five (four MLPs in the first stage and one MLP in the second stage). }
\end{figure*}

On the other hand, the performance of second-order training clearly surpasses that of first-order training.  Figure~\ref{fig:LM_vs_SD} shows a performance comparison between first- and second-order training: second-order training converges in a few hundred iterations for a simple illustrative curve-fitting task, whereas first-order training is not yet converged even after 25,000 iterations.   Thus, we conclude that second-order training finds a better path to a good solution compared to first-order methods.

\subsubsection{Network parameters}

Unless stated otherwise, a 3-stage MST system with the following configuration was used in the experiments herein: Stage 1: 60 MLPs, 2 hidden layers/MLP, 10 neurons/layer.  Stage 2: 30 MLPs, 2 hidden layer/MLP, 15 neurons/layer.  Stage 3: 30 MLPs, 2 hidden layer/MLP, 15 neurons/layer.  The details of the MST experiment are provided in Table~\ref{MSTtable}.   MST was implemented in MATLAB using \texttt{tansig} (hyperbolic tangent) activation functions~\cite{matlab:tansig} in the hidden layers and \texttt{purelin} (linear) in the output layers~\cite{matlab:purelin}.  

\begin{table*}
\centering
\begin{tabular}{lllllll}
\hdcell{}         & \multicolumn{2}{c}{\hdcell{Stage1}}                                                                                                                                                                             & \multicolumn{2}{c}{\hdcell{Stage2}}                                                                                                                                                                             & \multicolumn{2}{c}{\hdcell{Stage3}}                                                                                                                                                                             \\ 
\textbf{Settings} &              & \begin{tabular}[c]{@{}l@{}}Hidden layers per MLP: 2\\ Neurons per layer: 10\\ Maximum iterations: 100\\ Additional stopping criteria: \\ Mean squared error = 10$^{-3}$\end{tabular} &              & \begin{tabular}[c]{@{}l@{}}Hidden layers per MLP: 2\\ Neurons per layer: 15\\ Maximum iterations: 150\\ Additional stopping criteria: \\ Mean squared error = 10$^{-5}$\end{tabular} &              & \begin{tabular}[c]{@{}l@{}}Hidden layers per MLP: 2\\ Neurons per layer: 15\\ Maximum iterations: 250\\ Additional stopping criteria: \\ Mean squared error = 10$^{-7}$\end{tabular} \\
\hdcell{Layers}   & \hdcell{MLP} & \hdcell{Target}                                                                                                                                                                                  & \hdcell{MLP} & \hdcell{Target}                                                                                                                                                                                  & \hdcell{MLP} & \hdcell{Target}                                                                                                                                                                                  \\
                  & 1S1          & 1 for Tx1, 0 otherwise                                                                                                                                                                           & 1S2          & 1 for Tx1, 0 otherwise                                                                                                                                                                           & 1S3          & 1 to 12 for Tx1 to Tx12                                                                                                                                                                          \\
\rowcolor[HTML]{EFEFEF} 
                  & 2S1          & 1 for Tx1, 0 otherwise                                                                                                                                                                           & 2S2          & 1 for Tx2, 0 otherwise                                                                                                                                                                           & 2S3          & 1 to 12 for Tx1 to Tx12                                                                                                                                                                          \\
                  & 3S1          & 1 for Tx2, 0 otherwise                                                                                                                                                                           & 3S2          & 1 for Tx3, 0 otherwise                                                                                                                                                                           & 3S3          & 1 to 12 for Tx1 to Tx12                                                                                                                                                                          \\
\rowcolor[HTML]{EFEFEF} 
                  & 4S1          & 1 for Tx2, 0 otherwise                                                                                                                                                                           & 4S2          & 1 for Tx4, 0 otherwise                                                                                                                                                                           & 4S3          & 1 to 12 for Tx1 to Tx12                                                                                                                                                                          \\
                  & 5S1          & 1 for Tx3, 0 otherwise                                                                                                                                                                           & 5S2          & 1 for Tx5, 0 otherwise                                                                                                                                                                           & 5S3          & 1 to 12 for Tx1 to Tx12                                                                                                                                                                          \\
\rowcolor[HTML]{EFEFEF} 
                  & 6S1          & 1 for Tx3, 0 otherwise                                                                                                                                                                           & 6S2          & 1 for Tx6, 0 otherwise                                                                                                                                                                           & 6S3          & 1 to 12 for Tx1 to Tx12                                                                                                                                                                          \\
                  & 7S1          & 1 for Tx4, 0 otherwise                                                                                                                                                                           & 7S2          & 1 for Tx7, 0 otherwise                                                                                                                                                                           & 7S3          & 1 to 12 for Tx1 to Tx12                                                                                                                                                                          \\
\rowcolor[HTML]{EFEFEF} 
                  & 8S1          & 1 for Tx4, 0 otherwise                                                                                                                                                                           & 8S2          & 1 for Tx8, 0 otherwise                                                                                                                                                                           & 8S3          & 1 to 12 for Tx1 to Tx12                                                                                                                                                                          \\
                  & 9S1          & 1 for Tx5, 0 otherwise                                                                                                                                                                           & 9S2          & 1 for Tx9, 0 otherwise                                                                                                                                                                           & 9S3          & 1 to 12 for Tx1 to Tx12                                                                                                                                                                          \\
\rowcolor[HTML]{EFEFEF} 
                  & 10S1         & 1 for Tx5, 0 otherwise                                                                                                                                                                           & 10S2         & 1 for Tx10, 0 otherwise                                                                                                                                                                          & 10S3         & 1 to 12 for Tx1 to Tx12                                                                                                                                                                          \\
                  & 11S1         & 1 for Tx6, 0 otherwise                                                                                                                                                                           & 11S2         & 1 for Tx11, 0 otherwise                                                                                                                                                                          & 11S3         & 1 to 12 for Tx1 to Tx12                                                                                                                                                                          \\
\rowcolor[HTML]{EFEFEF} 
                  & 12S1         & 1 for Tx6, 0 otherwise                                                                                                                                                                           & 12S2         & 1 for Tx12, 0 otherwise                                                                                                                                                                          & 12S3         & 1 to 12 for Tx1 to Tx12                                                                                                                                                                          \\
                  & 13S1         & 1 for Tx7, 0 otherwise                                                                                                                                                                           &              &                                                                                                                                                                                                  &              &                                                                                                                                                                                                  \\
\rowcolor[HTML]{EFEFEF} 
                  & 14S1         & 1 for Tx7, 0 otherwise                                                                                                                                                                           &              &                                                                                                                                                                                                  &              &                                                                                                                                                                                                  \\
                  & 15S1         & 1 for Tx8, 0 otherwise                                                                                                                                                                           &              &                                                                                                                                                                                                  &              &                                                                                                                                                                                                  \\
\rowcolor[HTML]{EFEFEF} 
                  & 16S1         & 1 for Tx8, 0 otherwise                                                                                                                                                                           &              &                                                                                                                                                                                                  &              &                                                                                                                                                                                                  \\
                  & 17S1         & 1 for Tx9, 0 otherwise                                                                                                                                                                           &              &                                                                                                                                                                                                  &              &                                                                                                                                                                                                  \\
\rowcolor[HTML]{EFEFEF} 
                  & 18S1         & 1 for Tx9, 0 otherwise                                                                                                                                                                           &              &                                                                                                                                                                                                  &              &                                                                                                                                                                                                  \\
                  & 19S1         & 1 for Tx10, 0 otherwise                                                                                                                                                                          &              &                                                                                                                                                                                                  &              &                                                                                                                                                                                                  \\
\rowcolor[HTML]{EFEFEF} 
                  & 20S1         & 1 for Tx10, 0 otherwise                                                                                                                                                                          &              &                                                                                                                                                                                                  &              &                                                                                                                                                                                                  \\
                  & 21S1         & 1 for Tx11, 0 otherwise                                                                                                                                                                          &              &                                                                                                                                                                                                  &              &                                                                                                                                                                                                  \\
\rowcolor[HTML]{EFEFEF} 
                  & 22S1         & 1 for Tx11, 0 otherwise                                                                                                                                                                          &              &                                                                                                                                                                                                  &              &                                                                                                                                                                                                  \\
                  & 23S1         & 1 for Tx12, 0 otherwise                                                                                                                                                                          &              &                                                                                                                                                                                                  &              &                                                                                                                                                                                                  \\
\rowcolor[HTML]{EFEFEF} 
                  & 24S1         & 1 for Tx12, 0 otherwise                                                                                                                                                                          &              &                                                                                                                                                                                                  &              &                                                                                                                                                                                                 
\end{tabular}
\caption{MST method uses individual MLPs trained with the LM algorithm. The notation used for naming individual MLPs is mSs where m is the MLP number and s is the stage number. Only 18 MLPs out of 60 are shown in stage 1, 12 MLPs out of 30 are shown in stage 2 and 10 MLPs out of 30 are shown in stage 3 for compactness. Stopping criteria for MLPs in all stages include: the validation error not improving for 10 consecutive iterations, $\mu$ parameter in LM update rule is greater than $10^8$ for 10 consecutive iterations, mean square error reaching a certain threshold specified separately for each stage. The outputs shown for each stage show the desired response of each MLP to different transmitters. For example MLP 1S1 is trained to fire only if the input corresponds to transmitter 1, whereas groups of MLPs are trained to fire for different transmitters in stages 1 and 2. MLPs in stage 3 are trained to give a different response corresponding to the transmitter number. \label{MSTtable} }
\end{table*}

\subsubsection{Complexity advantage of MST}

Regardless of which method is used to compute weights update,  MST alone offers important advantages over conventional ML algorithms because of reduced computational complexity arising from the way in which MST-based training is done. This reduced complexity enables the use of second-order training algorithms on a much larger scale than typically possible. With second-order training, the main bottleneck is the computation of the Hessian matrix inverse.   In this context, MST improves computational efficiency in two ways.

The first way is by using multiple smaller matrices instead of a single large matrix for operations involving Jacobian and Hessian. Consider the system configuration used herein for RF signal identification as an example for an input size of 1,024 samples, the total number of parameters in the MST system is $N$=674,480 parameters (Table~\ref{tab:npar}).  Imagine a single MLP (such as CNN) with this many parameters.  Second-order training of such a single giant MLP would require inversion of a Hessian matrix of size 674,480$^2$, which would be exceedingly challenging from a computational standpoint.

\begin{table}
  {\footnotesize
\begin{tabular}{lc}
  \hdcell{MST stage} & \hdcell{Number of parameters} \\
  \hline
$1^{\mbox{\footnotesize\it st}}$ stage MLP & $1024\times 10 + 10\times 10 + 20 = 10,340$ \\
$2^{\mbox{\footnotesize\it nd}}$ stage MLP & $60\times 15 + 15 \times 15 + 30 = 1,145$ \\
$3^{\mbox{\footnotesize\it rd}}$ stage MLP & $30\times 15 + 15\times 15 + 30 = 705$ \\
\hline
Total & $60\times 10,340 + 30\times 1,145 + 30\times 705 = 674,480$ \\
\hline
\end{tabular}
}
\caption{Number of parameters in MST\label{tab:npar}}
\end{table}

In contrast, MST only requires the inversion of much smaller Hessian matrices. In the present study, MST  requires 60 Hessian matrices each with 10,360$^2$ elements ($1^{\mbox{\footnotesize\it st}}$ stage), 30 Hessians of size 1,145$^2$ ($2^{\mbox{\footnotesize\it nd}}$ stage), and 30 Hessians with size 705$^2$ ($3^{\mbox{\footnotesize\it rd}}$ stage).    If one uses the best matrix inversion algorithm~\cite{bib:matinv} available, which has complexity of $O(N^{2.373})$, MST would be 334 times faster per iteration, i.e. $674,480^{2.373}/(60\times10,360^{2.373} + 30 \times 1,145^{2.373} + 30\times 705^{2.373})$=334.

The second way an MST increases efficiency is by allowing parallel training of all MLPs at each stage. For the same example, we find that MST training would be 19,982 times faster than training one single giant MLP with the same number of parameters, given a full parallel implementation (e.g., using 60 parallel processing units), i.e. $674,480^{2.373}/(10,360^{2.373}+1,145^{2.373}+705^{2.373})$=19,982. 

This drastic improvement in computation time is also accompanied by a drastic reduction in storage memory requirements.  For a non-parallel implementation, our example MST requires 4,168 times less memory for storing the Hessian, i.e. $674,480^2/(10,360^2+1,145^2+705^2)$=4,168.   A parallel-processing implementation of MST (outside the scope of this study) would consume 70 times less memory, i.e. $674,480^2/(60\times 10,360^2 + 30\times 1,145^2 + 30\times 705^2)$=70.

\subsubsection{First-Order Training Analysis \label{sec:first-vs-second}\label{sec:first-order-mst}}

In this section we examine the question, {\bf Does the MST owe its performance  to the multi-stage training strategy, or to our use of second-order (LM)  training updates?} [Note that both CNN and DNN use a first-order update rule (stochastic gradient) during the back-propagation part of the training phase.]

Second-order order training via the LM algorithm~\cite{bib:yu} is known to get better results than first-order methods in fewer iterations~\cite{bib:yu,bib:huang06,bib:saravanan,bib:hagen94,bib:hertz91,bib:battiti92}. The MATLAB documentation states, ``{\tt trainlm} is often the fastest back-propagation algorithm in the toolbox, and is highly recommended as a first-choice supervised algorithm, although it does require more memory than other algorithms.''~\cite{matlab:lm}

The A-LM algorithm extends the applicability of second-order methods to large scale ANNs.  In order to demonstrate the power of second-order training, we compared the performance of MST under conditions of first- and second-order training.
The results (Table~\ref{tab:table_1storder}) show that while the performance of MST with second-order training was superior in terms of accuracy (as expected), the execution time was also faster than first-order training. This is due to the fact that while a single iteration of first-order training can be faster than a second-order training iteration, convergence requires substantially more iterations.

\begin{table}
\centering
\begin{tabular}{ p{1.8cm} p{1.61cm} p{1.61cm} p{1.61cm} p{1.61cm} }
\hdcell{Method}        & \hdcell{MST-1 $1^{\mbox{\footnotesize\it st}}$ order}  & \hdcell{MST-2 $1^{\mbox{\footnotesize\it st}}$ order} & \hdcell{MST-1  $2^{\mbox{\footnotesize\it nd}}$ order} & \hdcell{MST-2  $2^{\mbox{\footnotesize\it nd}}$ order}  \\ 
\hdcell{Accuracy}      & 91.35\%                       & 94.61\%                      & 96.8\%                        & 98.04\%                   \\ 
\hdcell{Training Time (rel.)} & 1.8                   & 6.9                   & 1.0                      & 5.3                \\ 
\end{tabular}
\caption{\label{tab:table_1storder} Analysis of MST performance under first- and second-order training.  MST-1 refers to the standard configuration used for previous experiments (60-30-30). MST-2 refers to a configuration with 3 times the number of MLPs at each stage. For the same experiment on the same data set, DNN had accuracy of $84.8\%$ whereas 2 CNN + 1 FC had accuracy of $67.3\%$.   The dataset w32, 10/90 was used here.  Training times are given in arbitrary (relative) units.   }
\end{table}

Increasing the system complexity by tripling the number of MLPs at each stage yielded a significant enhancement in performance. This led us to the conclusion that it is possible to achieve high performance with MST under first-order training. However, in order to reach a performance that is comparable to second-order-trained MST, the system complexity needs to be increased significantly, to the point where first-order training loses its computational efficiency advantage.

\section{Results}

We conducted experiments to demonstrate the applicability of our method to identify unknown transmitters, using training from a subset of the available data from twelve transmitters. Results demonstrate the ability for classification, scalability and recognition of rogue/unknown signals.

\subsection{Basic Classification \label{sec:classification}}
  
\begin{figure}[tb]
  \begin{tabular}{cc}
    \includegraphics[width=0.495\columnwidth]{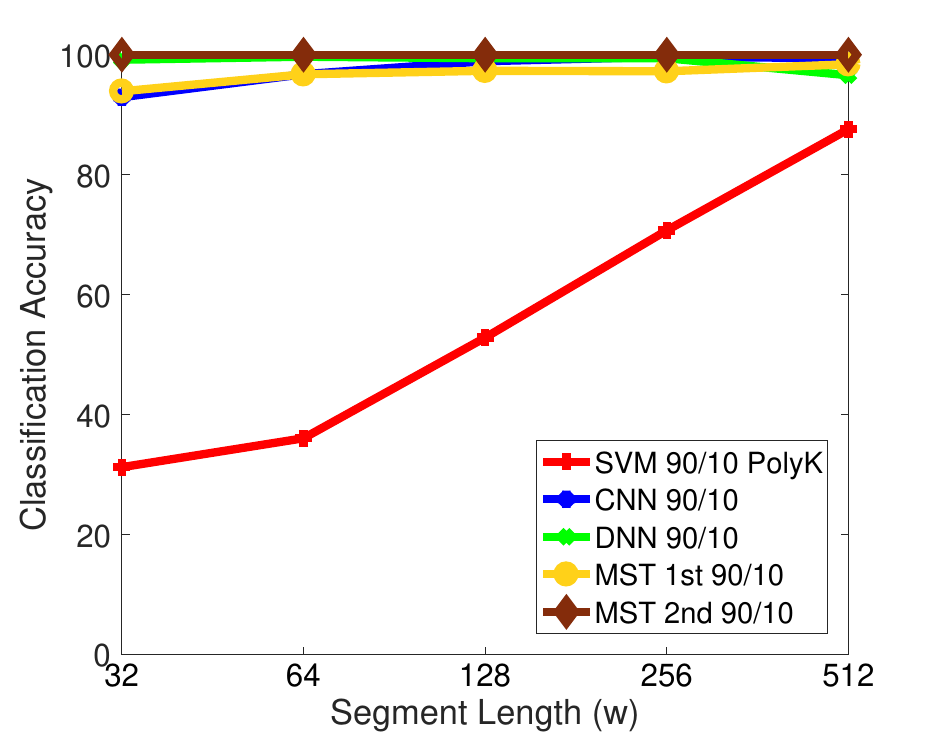}&
    \includegraphics[width=0.495\columnwidth]{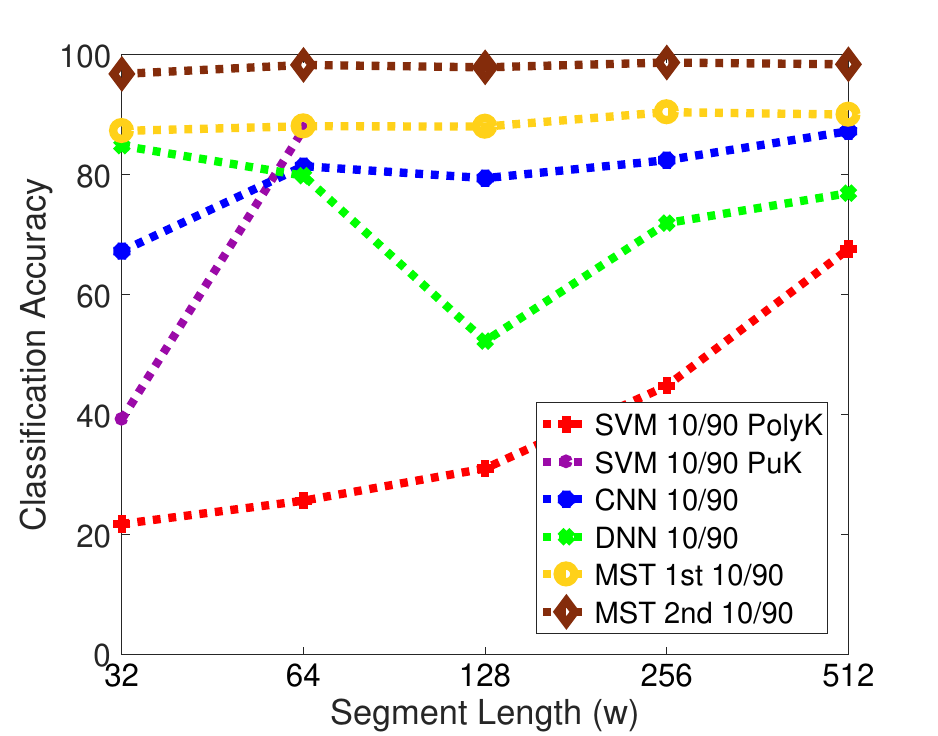} \\
  \end{tabular}
  \caption{ \label{fig:comp_methods} Comparison of the six algorithms:  MST
    clearly outperforms the other methods, particularly when trained on a much smaller dataset.
    Solid lines represent 90\% training, dotted lines represent 10\% training.
    Values appear in \tabref{table_L1a}.
  }
\end{figure}

\begin{figure}[tb]
  \begin{tabular}{cc}
    \includegraphics[width=0.495\columnwidth]{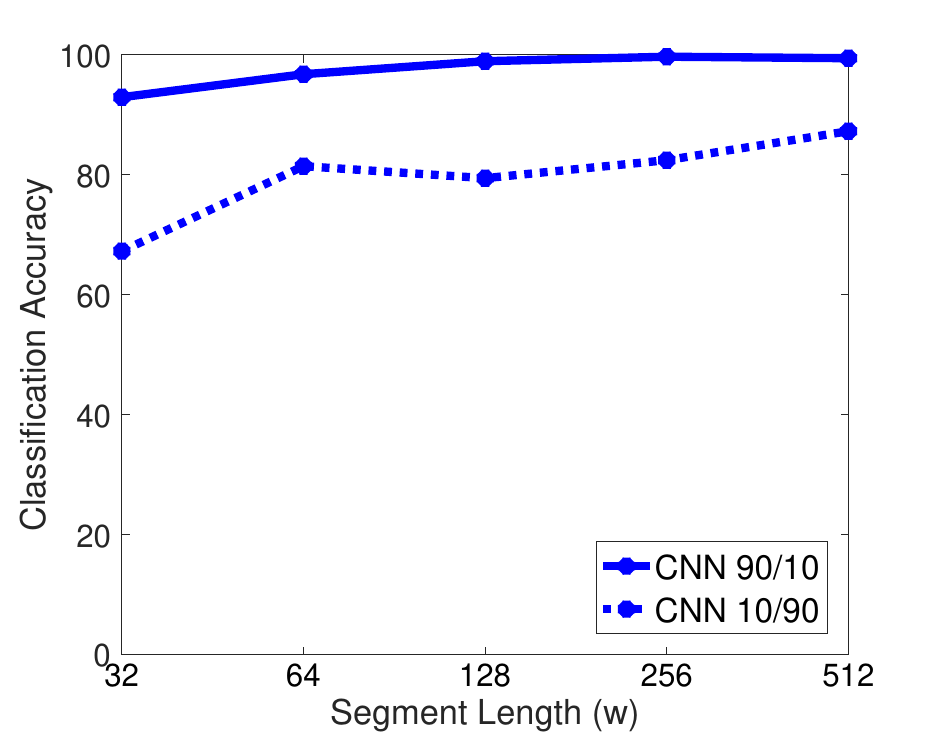} &
    \includegraphics[width=0.495\columnwidth]{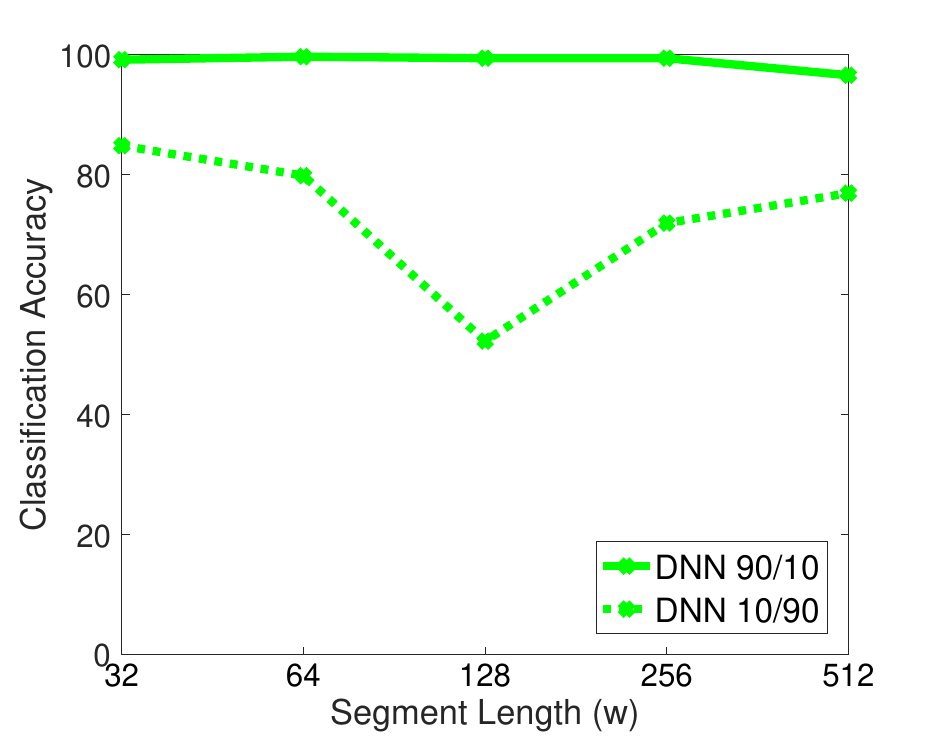} \\
    \includegraphics[width=0.495\columnwidth]{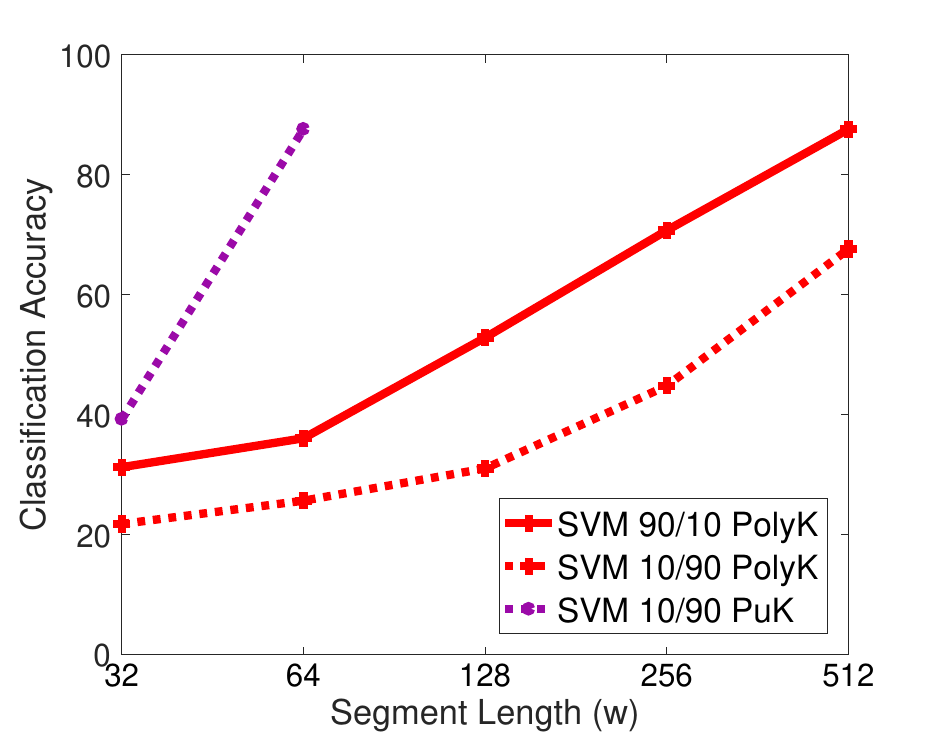} &
    \includegraphics[width=0.495\columnwidth]{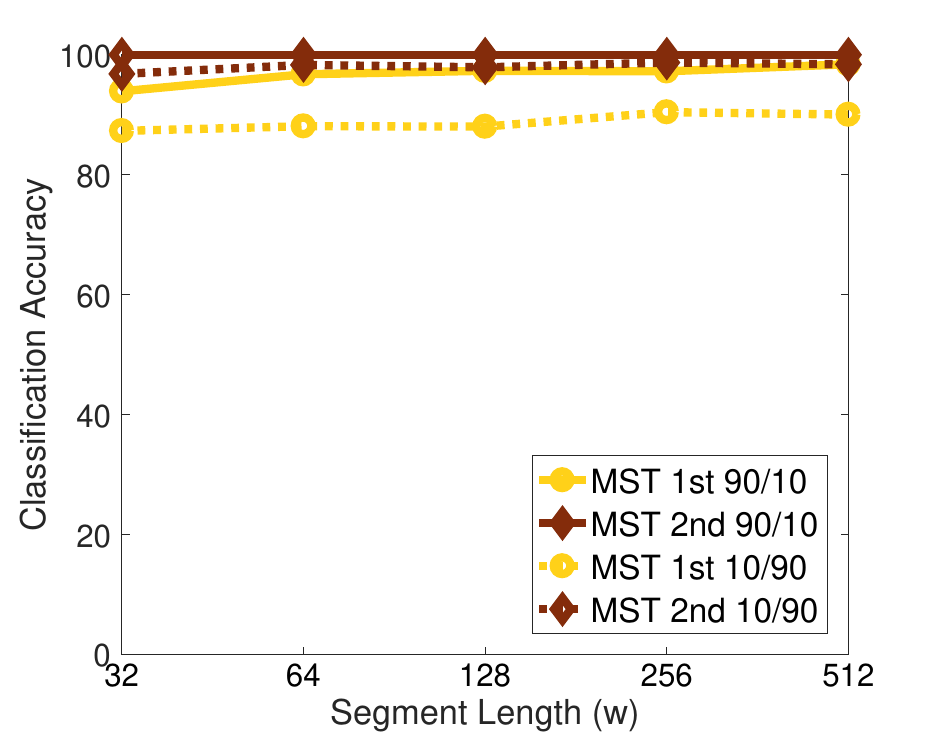} \\
  \end{tabular}
  \caption{ \label{fig:comp_methods_separated} Only MST maintains classification accuracy when the size of the training data is reduced. Note that SVM PuK ran out of memory for the larger datasets ($w  \geq 128$ for 10\% of the data, and all cases for 90\% training data.)  Values appear in \tabref{table_L1a}.
}
\end{figure}

{\bf In this section, we test the ability to accurately distinguish between a number of known transmitters.} Training was conducted using a percentage of the signals from the twelve transmitters (12,000 signals total). Given a new signal (not used in the training phase), the task consisted of identifying which transmitter it belongs to.  Table~\ref{tab:table_L1a}, Figures~\ref{fig:comp_methods} and~\ref{fig:comp_methods_separated} compare MST, CNN, DNN and SVM methods where 10\% or 90\% of the data were used for training. The remaining signals that were not used for training were used for testing.  The second-order trained A-LM MST method performed better under all conditions, and remained highly accurate even when trained using far less data (10/90).

Table~\ref{tab:table_L1a} also includes a comparison of first- and second-order trained MST performance.  For larger $N$ values (in w$N$), first-order training did not converge in reasonable time using the same MST configuration designed for second-order training.  Hence, a separate MST configuration optimized for first-order training was used for these comparisons.  The new configuration takes into account the inferior convergence properties of first-order training.   It spreads the desired cost function minimum goal into more stages, with more achievable intermediate goals at each stage.

A six-stage MST was used for first order training evaluation. Individual MLPs were trained using a gradient descent algorithm.  Stopping criteria for MLPs in all stages included: 1) the validation error not improving for 20 consecutive iterations, 2) mean square error reaching a certain number specified separately for each stage, 3) maximum number of iterations is reached (15,000 iterations). A large mean square error value was specified for the first stage ($10^{-1}$ as compared to $10^{-3}$ for 2$^{nd}$ order), and the goal was  slowly decreased over more stages (6 stages as compared to 3 stages for 2$^{nd}$ order) in order to compensate for the slow convergence of 1$^{st}$ order training, especially in the first stage, which is the most computationally demanding when the input size is large (e.g., w1024).  MLPs in stages one to five were trained to fire only if the input corresponded to a specific transmitter, where groups of MLPs were trained to fire for different transmitters. MLPs in stage 6 were trained to give a different response corresponding to the transmitter number.  The end result is that the second-order trained A-LM MST method outperformed first-order training under all conditions.

\begin{table}[tb]
  \noindent
  \footnotesize{
\begin{tabular}{lr rrrrrr}
\hdcell{Dataset} & \hdcell{Train \%} &  \hdcell{w32} & \hdcell{w64} & \hdcell{w128} & \hdcell{w256} & \hdcell{w512} & \\\hline
SVM PolyK & 90 &  31.2 & 36.0 & 52.8 & 70.7 & 87.6 & \\
SVM PuK & 90 &  NaN & NaN & NaN & NaN & NaN & \\
2 CNN + 1 FC & 90 &  92.9 & 96.8 & 98.9 & 99.7 & 99.4 & \\
DNN & 90 &  99.2 & 99.7 & 99.4 & 99.4 & 96.6 & \\
MST 1st order & 90 &  93.9 & 96.7 & 97.3 & 97.2 & 98.4 & \\
MST 2nd order & 90 &  100.0 & 100.0 & 100.0 & 100.0 & 100.0 & \\
\hline
SVM PolyK & 10 &  21.8 & 25.6 & 31.0 & 44.8 & 67.6 & \\
SVM PuK & 10 &  39.2 & 87.6 & NaN & NaN & NaN & \\
2 CNN + 1 FC & 10 &  67.3 & 81.4 & 79.4 & 82.4 & 87.3 & \\
DNN & 10 &  84.8 & 79.8 & 52.3 & 71.9 & 76.9 & \\
MST 1st order & 10 &  87.3 & 88.1 & 88.0 & 90.4 & 90.0 & \\
MST 2nd order & 10 &  96.8 & 98.3 & 97.9 & 98.7 & 98.4 & \\
\hline
\end{tabular}
}
\caption{ \label{tab:table_L1a} MST (second-order) outperformed (i.e. 100\% accuracy) all five other  algorithms, when trained with 90\% of the data, and dramatically outperformed  the other techniques when trained with only 10\% of the data.  (With the  larger datasets, Weka's  implementation of SVM PuK  ran out of memory before successfully building a model.)  These results  are plotted in Figures~\ref{fig:comp_methods}  and~\ref{fig:comp_methods_separated}.  }
\end{table}

In most cases, the SVM method underperformed relative to other methods. As expected, the SVM PuK kernel obtained markedly better results, but Weka's implementation of PuK is so inefficient that it ran out of memory before building a model for the larger datasets.

There is a contradictory effect of the segment length on the performance between the DNN and CNN systems. In the former case, the performance decreases as the length of the segments grows while the opposite effect is observed with the latter.  Our reasoning is that various artifacts will affect the signal in increasing number of combinations with the growing length and the DNN model is not robust enough to account for this variability. The CNN model applies filters locally and also incorporates the pooling mechanism, which we believe make it more robust. Also, the longer  the length of the input segments, the more device-specific patterns can be learned by the convolutional filters. Finally, the CNN model has more parameters and requires more data to achieve good performance, which explains the worse performance for the short segments.

The performance of both DNN and CNN models degrade significantly under the condition of limited  training data.  Our contrastive experiments also show that DNN training with limited amounts of data was much less stable in terms of the resulting accuracy, as demonstrated by the accuracy drop for the model trained for 128 samples long (w128) segments in Fig.~\ref{fig:comp_methods}.

\begin{figure*}
  \fbox{
  \begin{minipage}{0.95\textwidth}
      \begin{tabular}{ p{0.285\textwidth} p{0.285\textwidth} p{0.285\textwidth} p{0.09\textwidth}}
    \includegraphics[width=0.285\textwidth]{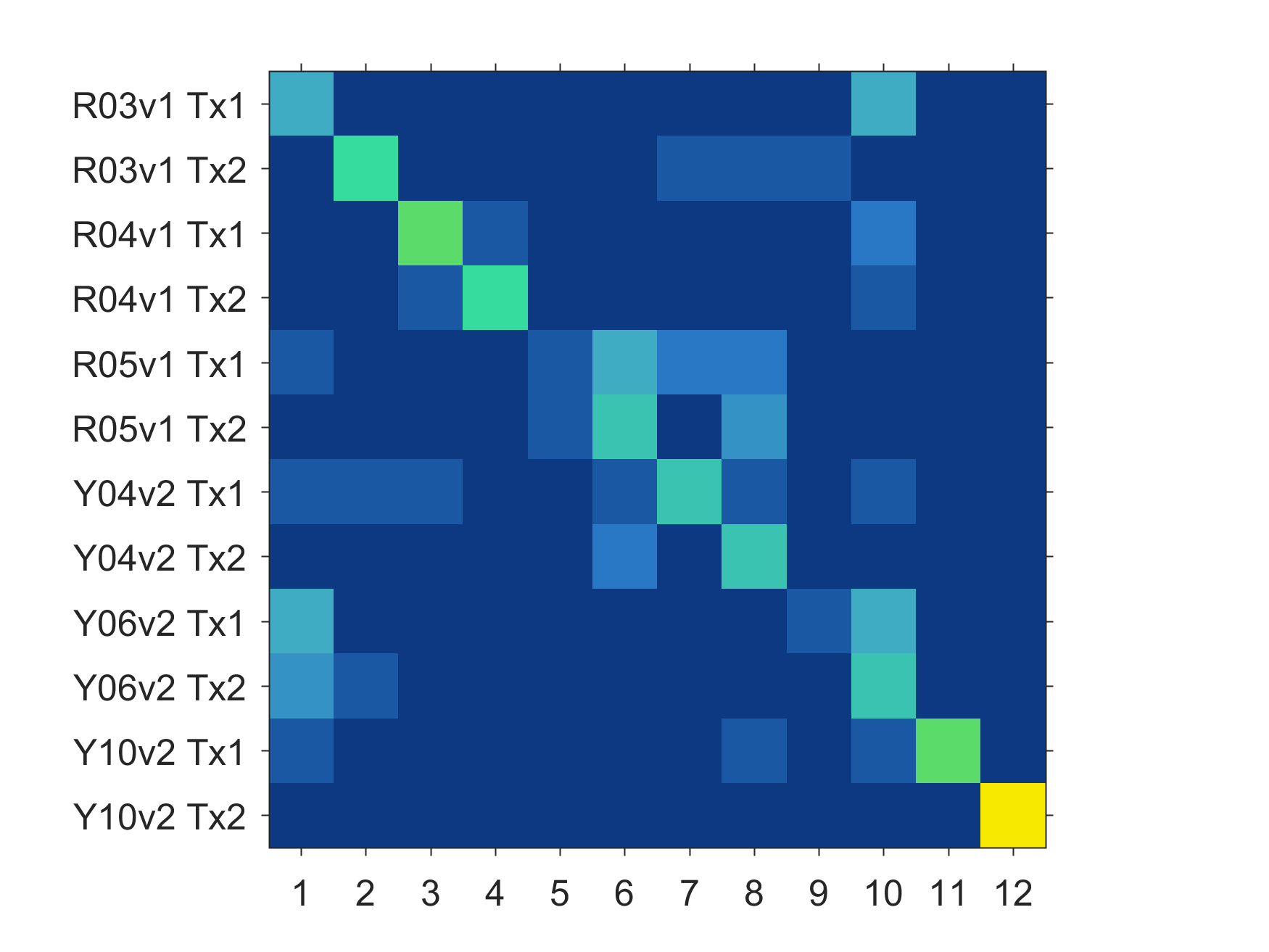} &
    \centerline{(Out of memory)} &
    \includegraphics[width=0.285\textwidth]{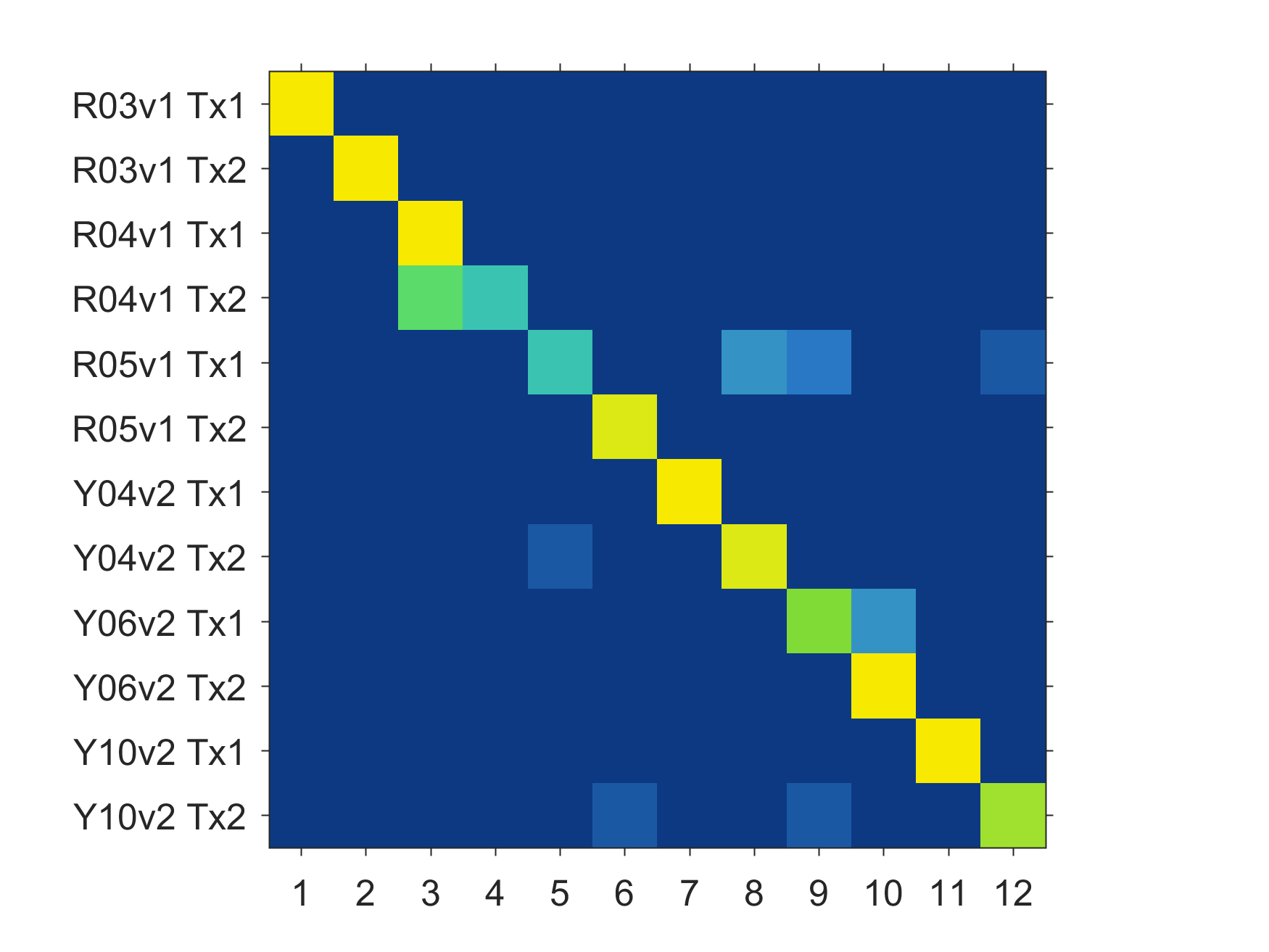} &
    \includegraphics[height=1.25in]{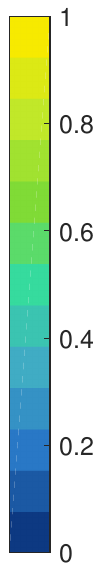}
    \\
    \centerline{(a) SVM, PolyKernel (44.8\%)} &
    \centerline{(b) SVM, PuK (N/A)} &
    \centerline{(c) CNN (82.4\%)} \\
    \includegraphics[width=0.285\textwidth]{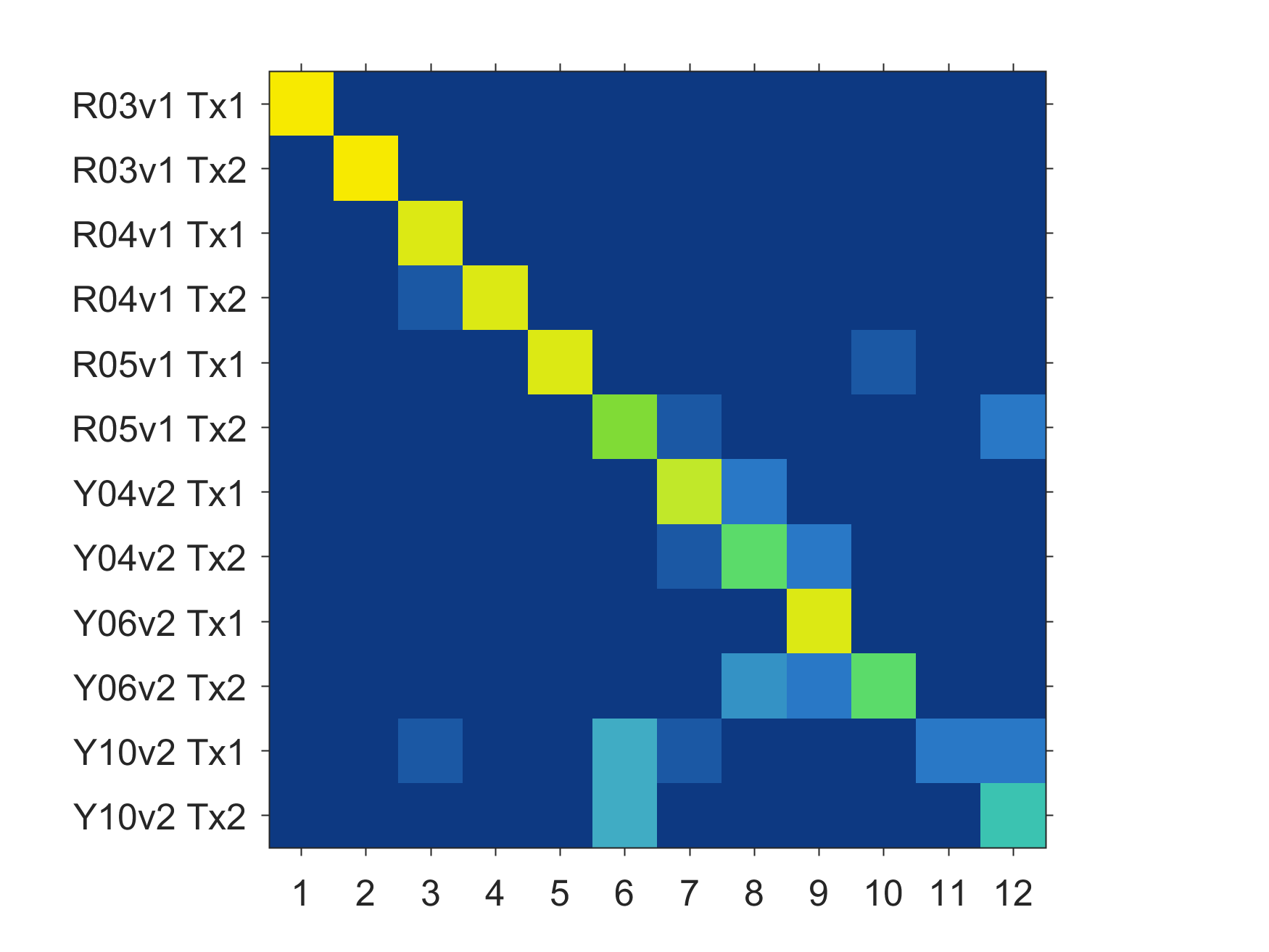}    &
    \includegraphics[width=0.285\textwidth]{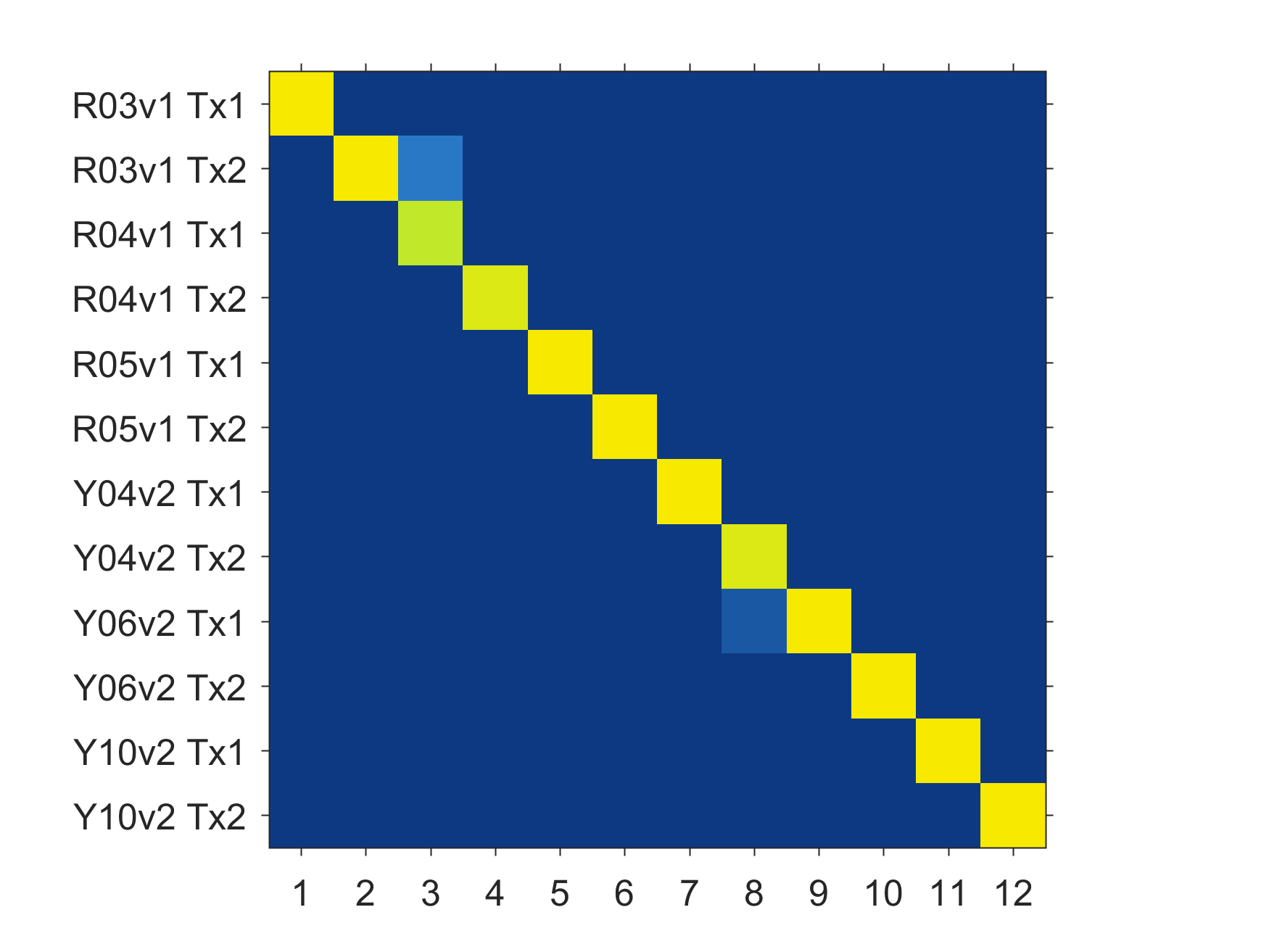}    &
    \includegraphics[width=0.285\textwidth]{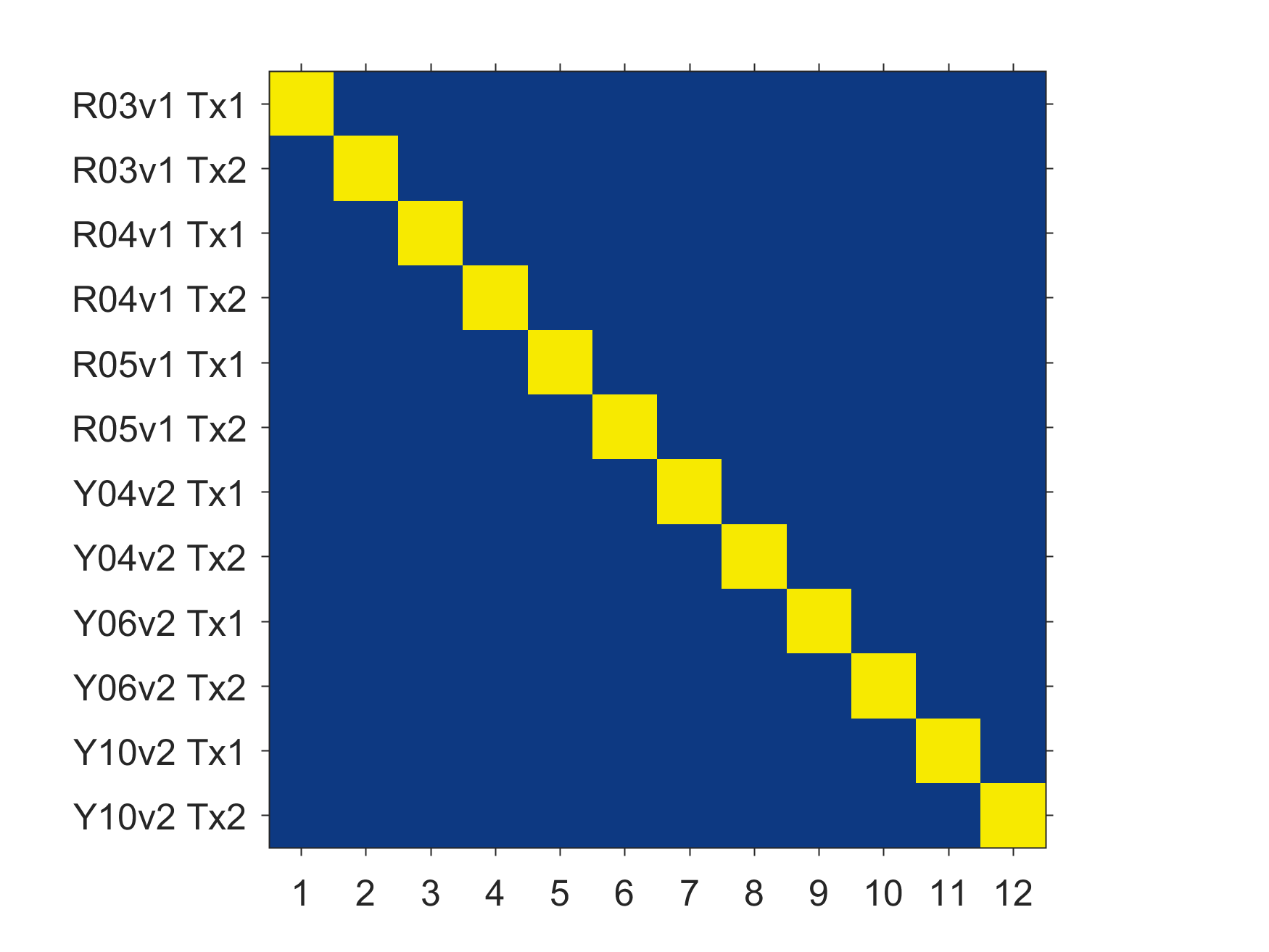}
    \\
    \centerline{(d) DNN (71.9\%)} &
    \centerline{(e) MST $1^{\mbox{\footnotesize\it st}}$ order (92.3\%)} & 
    \centerline{(f) MST  $2^{\mbox{\footnotesize\it nd}}$ order  (98.7\%)} 
  \end{tabular}
  \caption{\label{fig:confmat_256}
    These confusion matrices represent the
    labels for the 12 transmitters, 10/90 w256,  as 
      classified  for the six algorithms:
      SVM (PolyKernel and PuK \secpageref{svm}),
      DNN (\secpageref{dnn}),
      CNN (\secpageref{cnn}),
      MST (first-order \secpageref{first-order-mst} and second-order \secpageref{second-order-mst})
      w256 represents a relatively high-information state, and MST second-order
      achieves 98.7\% accuracy when trained on only 10\% of the data.
  }

  \end{minipage}
  }

  
  \fbox{
  \begin{minipage}{0.95\textwidth}
      \begin{tabular}{ p{0.285\textwidth} p{0.285\textwidth} p{0.285\textwidth} p{0.09\textwidth}}
          \includegraphics[width=0.285\textwidth]{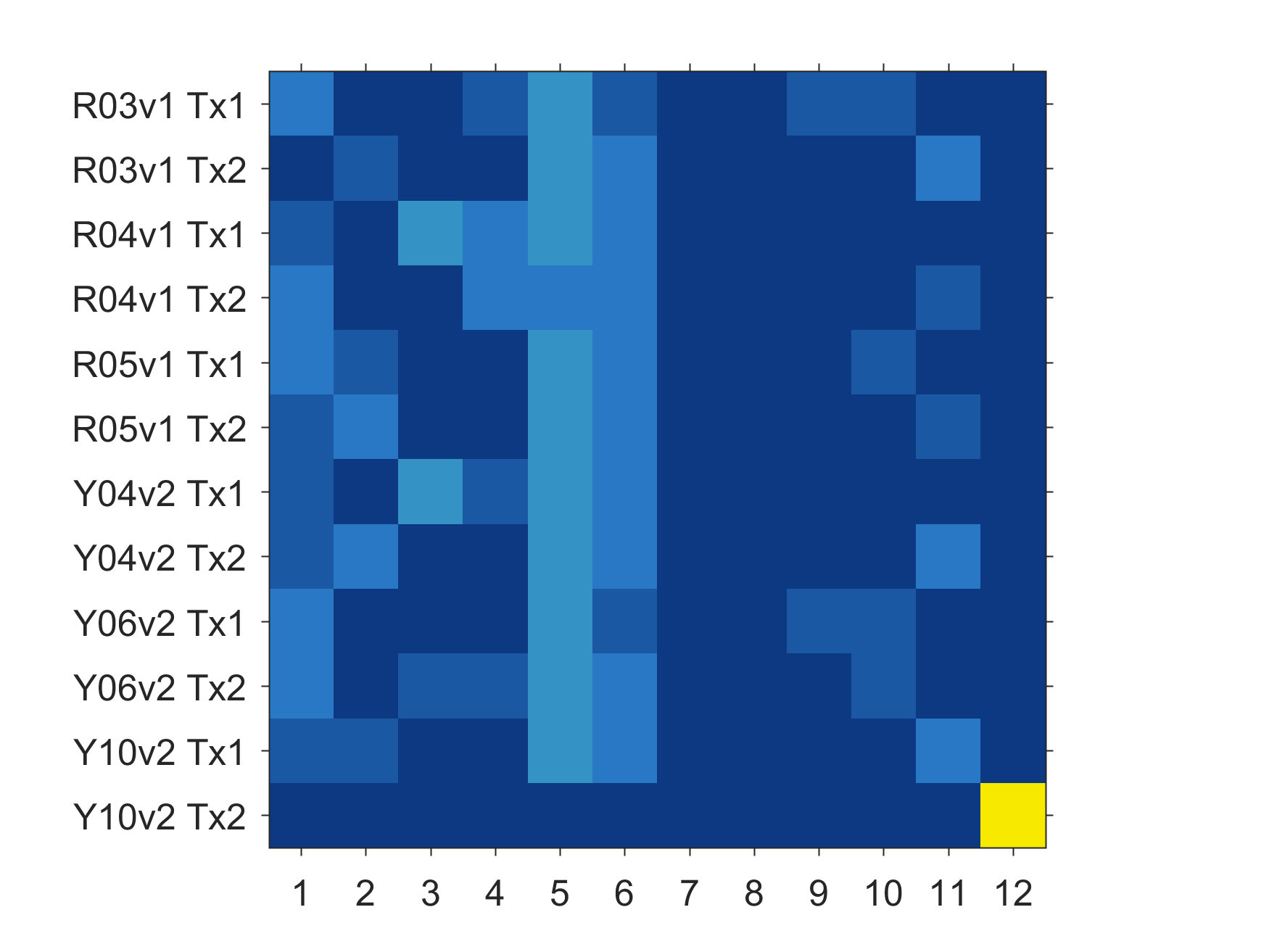} &
          \includegraphics[width=0.285\textwidth]{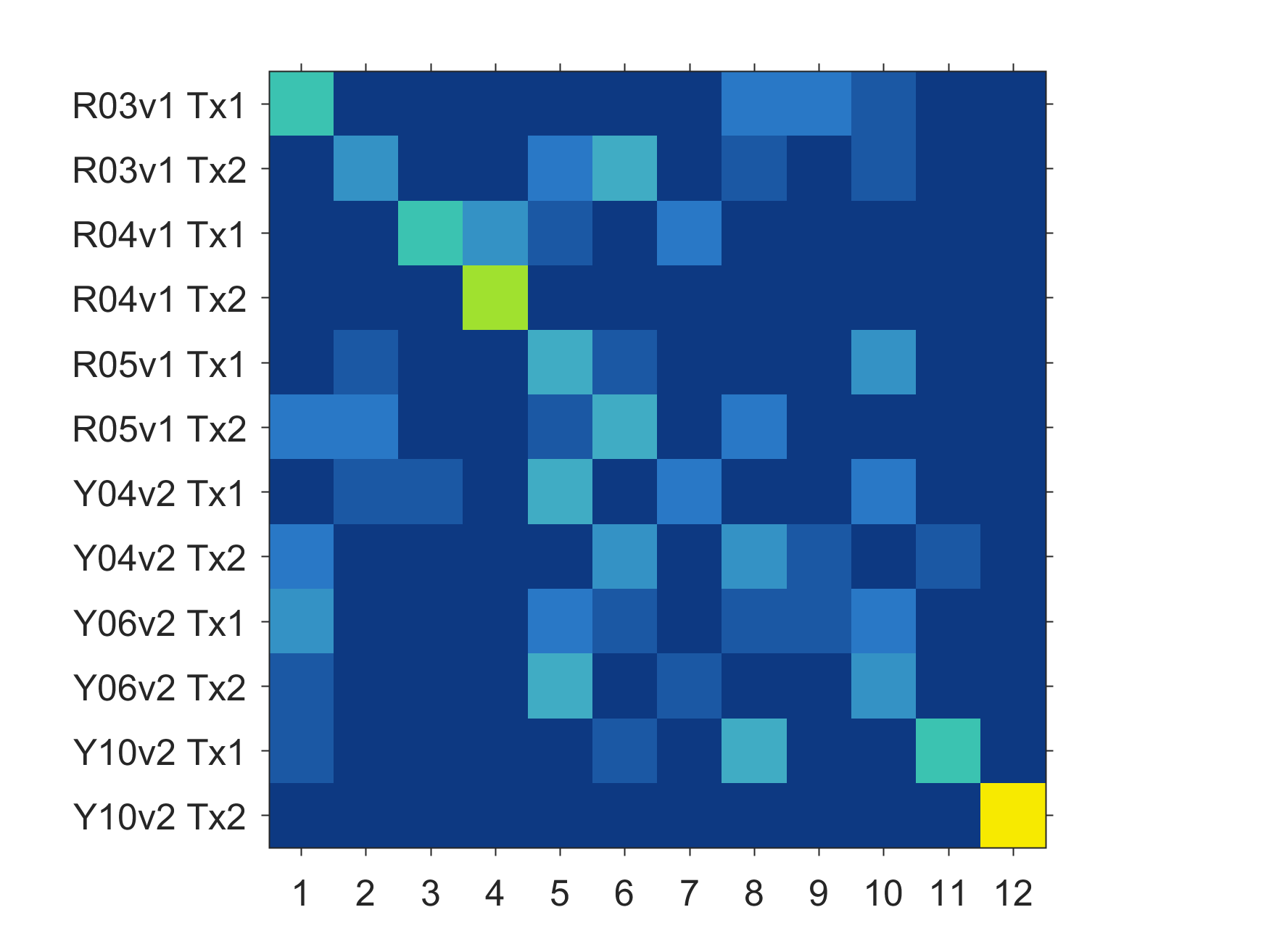} &
          \includegraphics[width=0.285\textwidth]{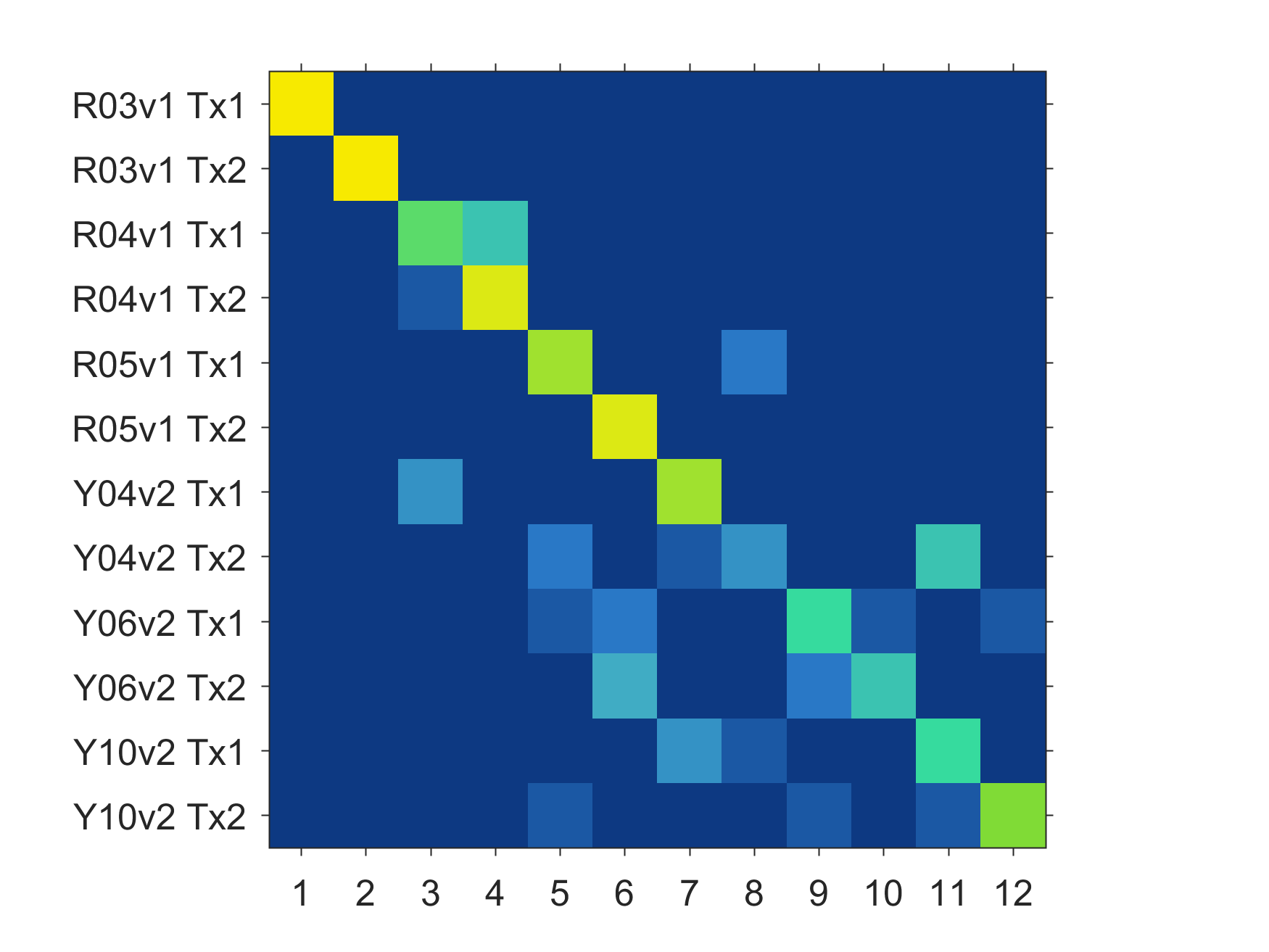} &
          \includegraphics[height=1.25in]{colourbar}
          \\
          \centerline{(a) SVM, PolyKernel (21.8\%)} &
          \centerline{(b) SVM, PuK (39.2\%)} &
          \centerline{(c) CNN       (67.3\%)} \\
          \includegraphics[width=0.285\textwidth]{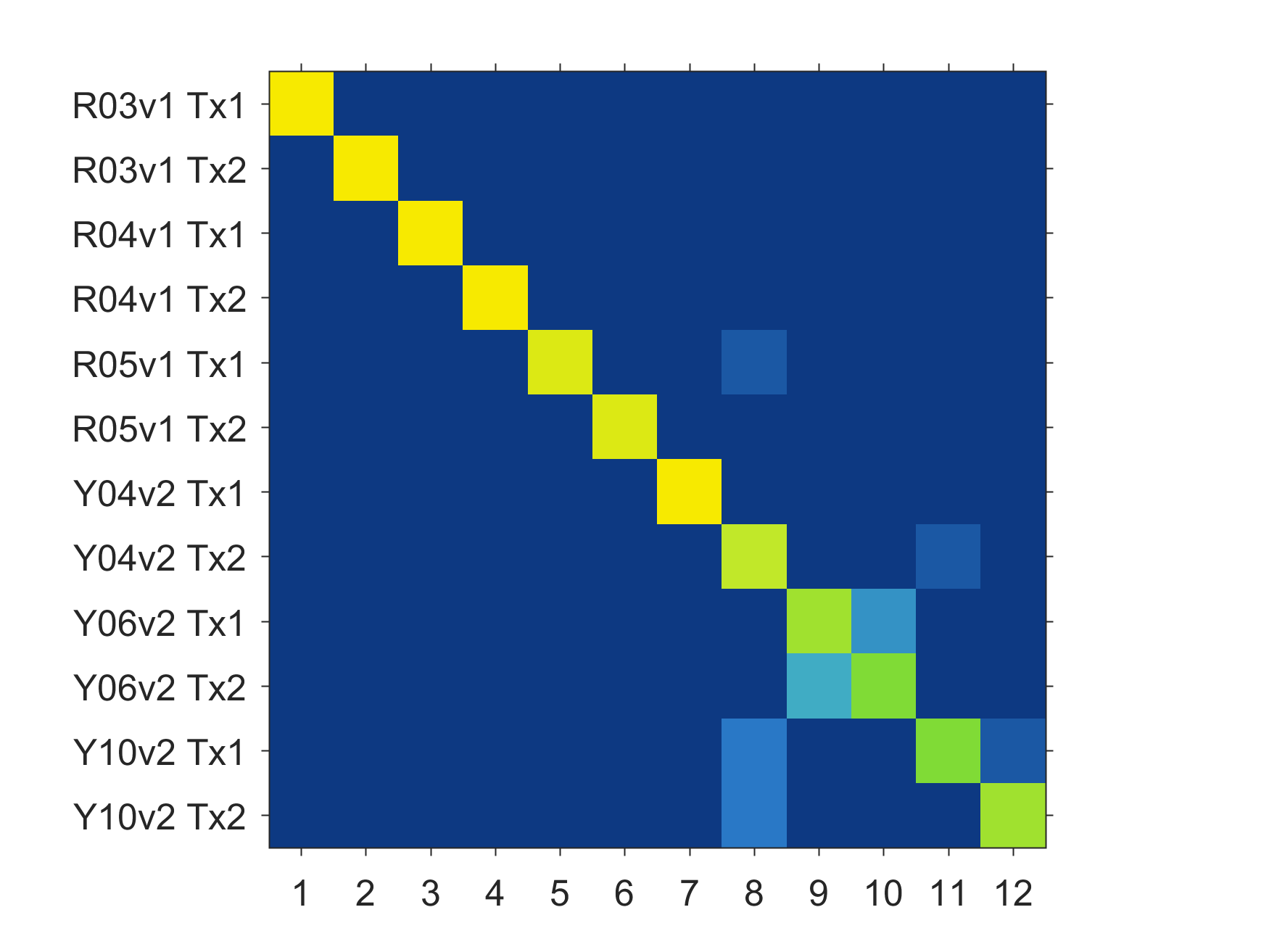}&
          \includegraphics[width=0.285\textwidth]{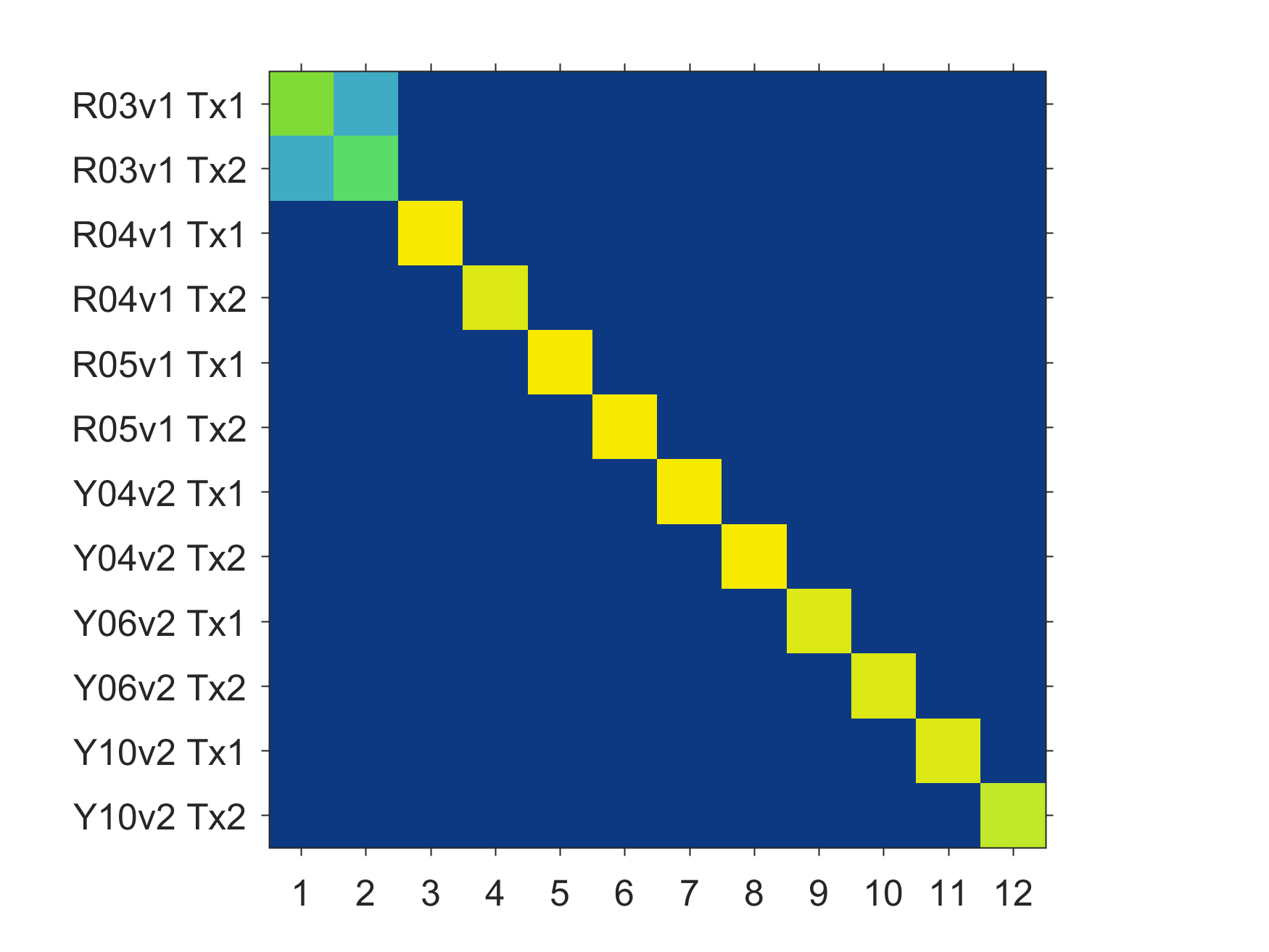} &
          \includegraphics[width=0.285\textwidth]{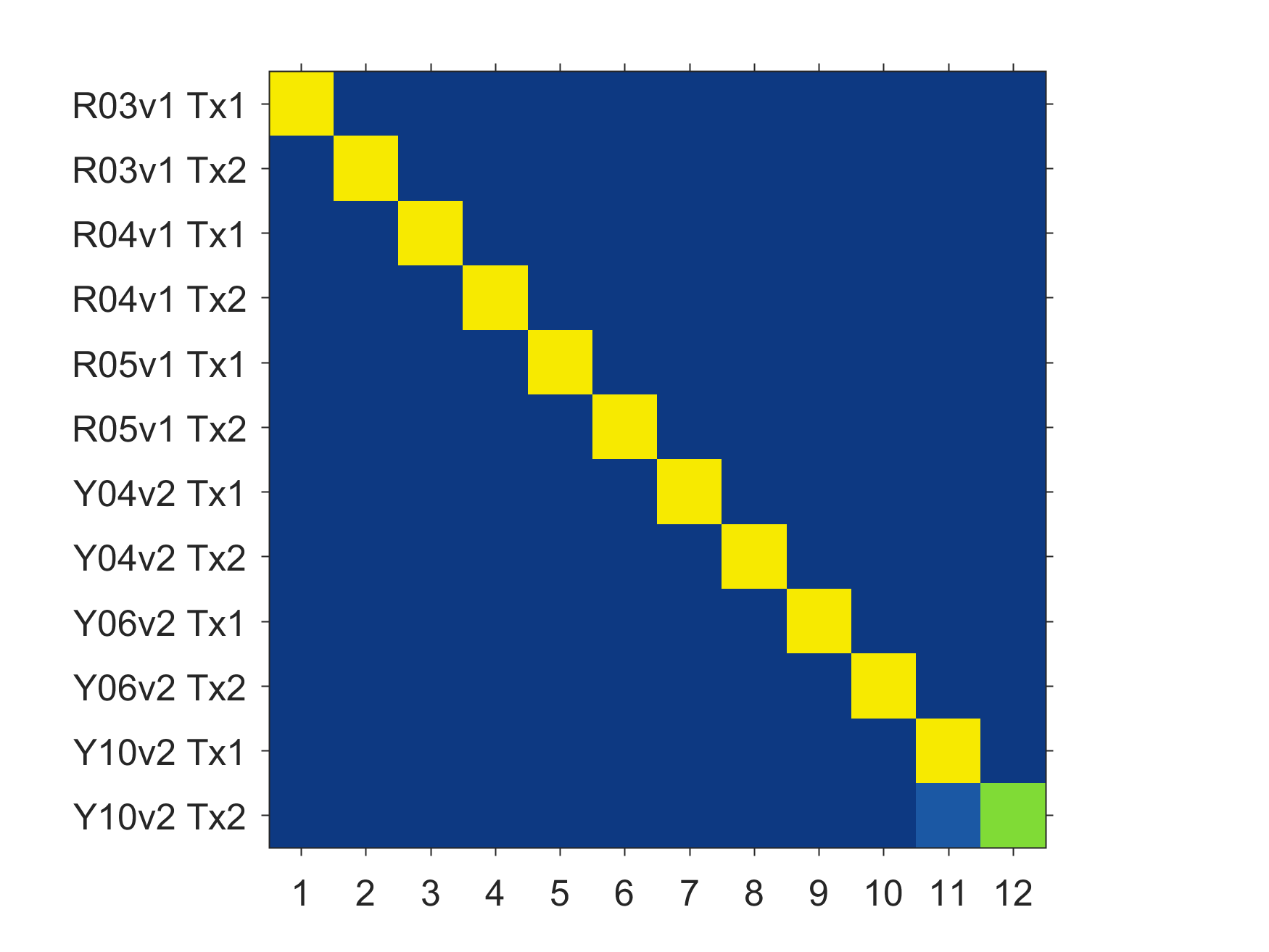}
          \\
          \centerline{(d) DNN  (84.8\%)} &
          \centerline{(e) MST $1^{\mbox{\footnotesize\it st}}$ order  (87.7\%)} & 
          \centerline{(f) MST  $2^{\mbox{\footnotesize\it nd}}$ order  (96.8\%)} \\
    \end{tabular}
    \caption{\label{fig:confMat_w32}
      These confusion matrices show the classification accuracy for the 12 transmitters,
      10/90 w32 (the hardest prediction to make), as
      classified  for the six algorithms:
      SVM (PolyKernel and PuK \secpageref{svm}),
      DNN (\secpageref{dnn}),
      CNN (\secpageref{cnn}),
      MST (first-order \secpageref{first-order-mst} and second-order \secpageref{second-order-mst})
      \textbf{Note} that confusions are  more likely between each pair of transmitter on
      the same radio (i.e. Tx1 and Tx2) than they are likely from one radio to
      another because because multiple transmitters on the same radio share a
      power supply and reference oscillator.
      MST, however, appears to be largely insensitive to this characteristic over the range of parameters investigated.
      \textbf{Note} also that Y10v2\_Tx2 is easy to identify due to its bad
      via, even for the weaker-performing algorithms.
    }

  \end{minipage}
  }
\end{figure*}

\figref{confmat_256} shows confusion matrices as obtained from all six models using {\bf w256 data} segments, and 10\% of the data for training. The A-LM MST method outperformed other methods.  This is rather surprising, given that w256 represents a relatively high-information model, and yet the SVM, CNN and DNN models were unable to achieve high accuracy.   In contrast, MST (2$^{nd}$ order) had nearly perfect accuracy (98.7\%), hence the appearance of an identity matrix.

\figref{confMat_w32} shows confusion matrices from all six models in a low-information mode, using only the first 32 samples of the signal ({w32}).  All methods (SVM, CNN, DNN) performed poorly, while MST (2$^{nd}$ order) maintained very high accuracy (96.8\%) in spite of the very limited training set.

\subsection{Incremental Learning \label{sec:incremental}}

Since the CNN and MST methods outperformed DNN and SVM in many of the classification tasks, we limited our further investigations to the CNN and MST methods. {\bf The question here is how easily can the model extend to capture new devices.}

We considered the scenario of extending a once trained and deployed system by the ability to recognize new devices. This task is typically referred to as incremental learning. To avoid building a unique classifier per device and enable low-shot learning of new devices, for CNN we used output nodes with sigmoid activations as used in multi-label classification. The advantage of this structure is that the front-end layers are shared across all device classifiers. In fact, each device detector differs only in the rows of the weight matrix between the last layer and the output layer.  Thus, adding a new device, which entails adding an extra output node, would simply require the estimate of one row of this weight matrix.

\subsubsection{CNN Model}

\begin{figure}[h]
\includegraphics[width=\columnwidth]{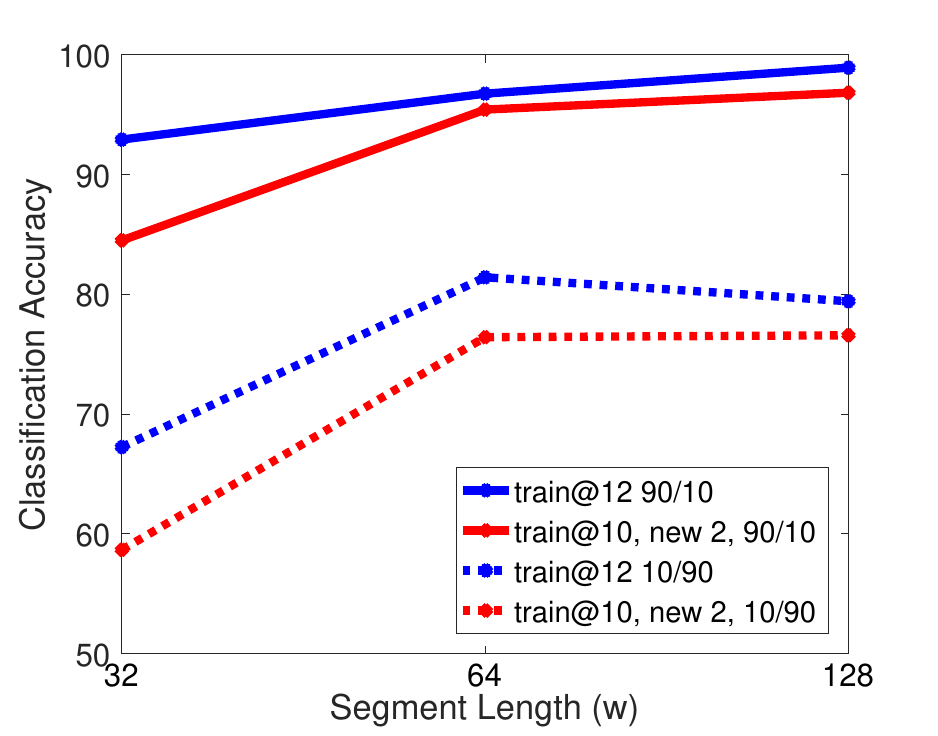}
\caption{ \label{fig:cnn_incrlearn} Incremental learning using the CNN
  multi-label approach achieves 98\% of the accuracy of comparable fully
  retrained model for $w\geq 64$.
}
\end{figure}

In our experimental setup, we fully train the original model with 10 devices. Another two new devices are then added to the model as described above. Figure~\ref{fig:cnn_incrlearn} compares the accuracy of two models. The first was trained on data from all 12 devices. The other one was trained on data from 10 devices and another two devices were registered with the model by means of extending the output layer. All hidden layers remain unaltered by the model extension. Contrastive experiments have shown that this technique works better with CNN models than DNN models, which demonstrates that the former generalize better as the representation extracted by the front-end layers have enough discriminative power to distinguish unseen devices. Another observation is that the performance drop for short segments is emphasized in this test condition. We attribute this to limited generalization of the set of device-specific patterns learned from the short segments.

\subsubsection{MST Model}

For the MST method the computational complexity of the system is largely determined by the number of MLPs in the first stage for two reasons. 1) The size of the input vector is typically larger than the number of MLPs in any stage, whereas increasing the number of transmitters may require increasing the size of the input vector. Thus, first stage MLPs typically have the highest number of parameters. 2) The number of MLPs in the first stage determines the number of inputs to the second stage. For example, in the previous experiment, we have used 5 MLPs for each transmitter in the first stage. Thus, increasing the number of transmitters will require training more MLPs in the first stage, which will also increase the size of the input to the second stage.

We designed the following experiment to test the ability of our method to use features learned from known transmitters to build representations for new transmitters.  In this classification task experiment, only data from $k$ out of $n$ transmitters was used to train the first stage. Data from all $n$ transmitters were then introduced in the second and third stages. The remarkable system performance, shown in Table~\ref{tab:table_L2}, even for the challenging case of w32, 10/90 training with $n/k=12/6$ (6 new devices), establishes MST (2$^{nd}$ order) as a much better alternative than CNN.  The ability to recognize 6 new devices knowing only 6 devices suggest that the MST ANN may possess the scalability property.  This scalability property will be critical for the expansion of the system to a larger number of transmitters, where only a small portion of transmitters would be needed to train the first stage.  This will dramatically reduce the complexity of the system.

\begin{table}[h]
\centering
\begin{tabular}{ p{1.5cm} p{1.30cm} p{1.30cm} p{1.30cm} p{1.30cm} }
  \hdcell{Training/ Testing} & \multicolumn{2}{c}{\hdcell{90/10}} &  \multicolumn{2}{c}{\hdcell{10/90}} \\
  \hdcell{n/k}               & \hdcell{12/6}    & \hdcell{12/11}  & \hdcell{12/6}    & \hdcell{12/11}   \\ 
  \textbf{w32}               & 98.83\%          & 99.75\%         & 97.47\%          & 98.76\%          \\ 
  \textbf{w64}               & 99.92\%          & 99.75\%         & 97.72\%          & 98.81\%          \\ 
  \textbf{w128}              & 99.75\%          & 100.00\%           & 97.94\%          & 98.91\%          \\ 
\end{tabular}
\caption{ \label{tab:table_L2} Ability of the MST (2$^{nd}$ order) method to identify new
  transmitters from a set of $n-k$ new transmitters when trained on data from
  only $k$ transmitters.
}
\end{table}

\section{Wavelet Preconditioning \label{sec:wprec} }

In this section we examine the question, {\bf Given the goal of identifying a very large number of unknown transmitters, can the performance of MST be further improved while keeping the network size relatively small?}  We propose a method of wavelet preconditioning to address this scalability problem.

The continuous wavelet transform (CWT) is the tool of choice for time-frequency analysis of 1D signals.  While no additional signal content is created, the CWT enables a clear depiction of the various frequency components of a signal as function of time.   Feeding the output of CWT to the ML module enhances the classification and identification tasks because the physical characteristics of the transmitter can be clearly separated in a two-dimensional representation. Furthermore, CWT allows representing the data in a much more compact way and reduces the number of parameters in the MST. For example, a time domain segment consisting of 2,048 samples that requires approximately 1.3 million parameters on the MST system given herein, can be efficiently represented by only 256 samples using CWT.  This reduces the number of parameters to 213,000. The drastic reduction in the number of parameters reduces system complexity while increasing convergence speed of the training algorithm.

\begin{figure}[h]
\includegraphics[width=\columnwidth]{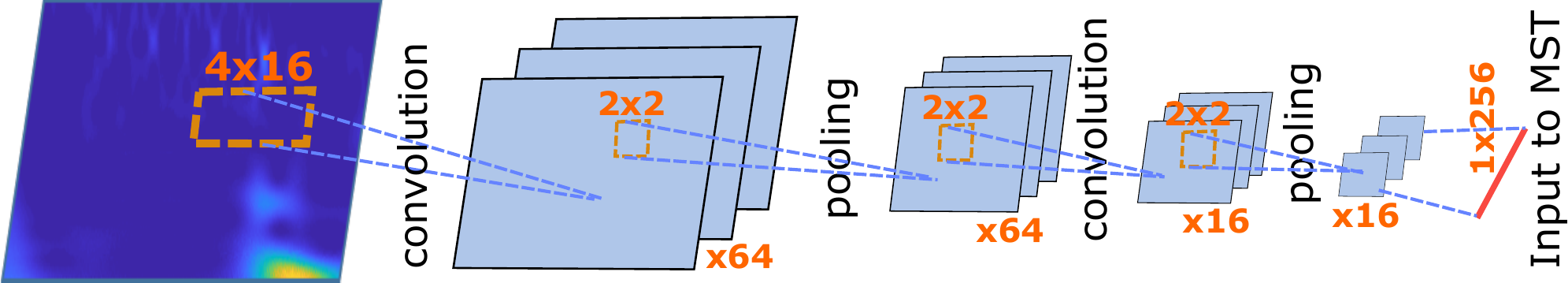}
\caption{ \label{fig:front_end} Wavelet preconditioning front-end is used to ``compress'' the data from an OFDM packet into a few key parameters, which are then fed to the RF machine learner.  This figure illustrates the architecture of the module used for RF signal feature extraction. The output of this module is sent to the MST stages for classification. Self-organizing map (SOM)~\cite{bib:CNNSOM}, an unsupervised ANN method, was used for selecting the filters weights in the convolution layers.}
\end{figure}

The continuous wavelet transform of a signal $f(t)$ is:
$$ (T^{wav} f)(a,b)=|a|^{-1/2} \int dt f(t) \overline{\psi}\left( \frac{ t-b}{a} \right). $$
$\psi(t)$ is a mother wavelet whose wavelets are:
$$ \psi^{a,b}(t) = |a|^{-1/2} \psi(\tfrac{t-b}{a}), $$
where $b\in\mathbb{R}$ is the translation, $a\in\mathbb{R}_+$ is the scale parameter with $a\ne 0$ and $\int d\xi |\xi|^{-1} | \hat{\psi}(\xi)|^2 < \infty$ ($\hat{\psi}$ is the Fourier transform of $\psi$). For our analysis, we picked the modulated Gaussian (Morlet) wavelet,
$$ \psi(t) = \pi^{-1/4} \left( e^{-i \xi_0 t } - e^{-\xi_0^2/2 } \right) e^{-t^2/2 } $$
and its MATLAB implementation in the command \texttt{cwt}, which produces wavelet scalograms that are then fed to a carefully designed ``wavelet preconditioning'' front-end (see Figure~\ref{fig:front_end} for the overall design).   Wavelet scalograms are fed to self-organizing map (SOM) and pooling layers~\cite{bib:CNNSOM} and used as a front end to the MST system to extract features from the wavelet transform and reduce the dimensionality of the input. 

\begin{figure}[h]
\includegraphics[width=\columnwidth]{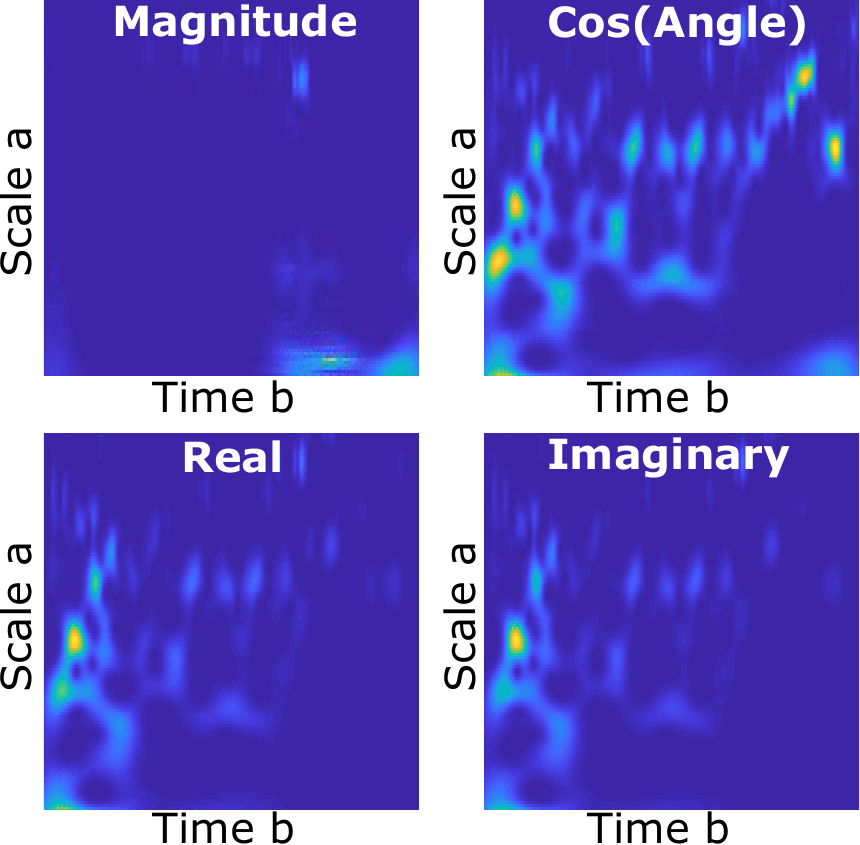}
\caption{ \label{fig:var_comp} Wavelet scalograms can be constructed by taking the CWT of the real, imaginary, phase (here, the cosine of the phase) or magnitude of the time-domain RF signal.  The CWT of the RF magnitude (top left) shows the least amount of features (sparsest representation).   This sparse representation yielded the best performance for classification.   }
\end{figure}

We performed a variance analysis of the wavelet transform scalogram across the various RF packets to gain insight about the information content of the packet.  In Figure~\ref{fig:var_comp} we plot the variance of the scalogram pixel intensities when taken across the multiple transmitters and signal packets.   Regions of high variance indicate which scalogram components may be more involved in the representation of RF signal features.    The OFDM waveforms $\vec{f}=(f_1,f_2,\dots,f_n)$ had $n=10,000$ data points.  We skipped the first 450 points and stored the next 2,048 points in a vector, i.e. $\vec{g}=(g_1,g_2,\dots,g_{2048})$, where $g_i=f_{i+450}$.  A vector $\vec{v}$ was constructed from the elements of $\vec{g}$ by taking either: 1) the absolute value, 2) the cosine of the phase angle, 3) the real part or 4) the imaginary parts of the complex-valued components, i.e. 1) $v_i=|g_i|$,  2) $v_i=\cos(\tan^{-1}(\Im(g_i)/\Re(g_i)))$, 3) $v_i=\Re(g_i)$ or 4) $v_i=\Im(g_i)$, respectively.

\begin{figure}[h]
\includegraphics[width=\columnwidth]{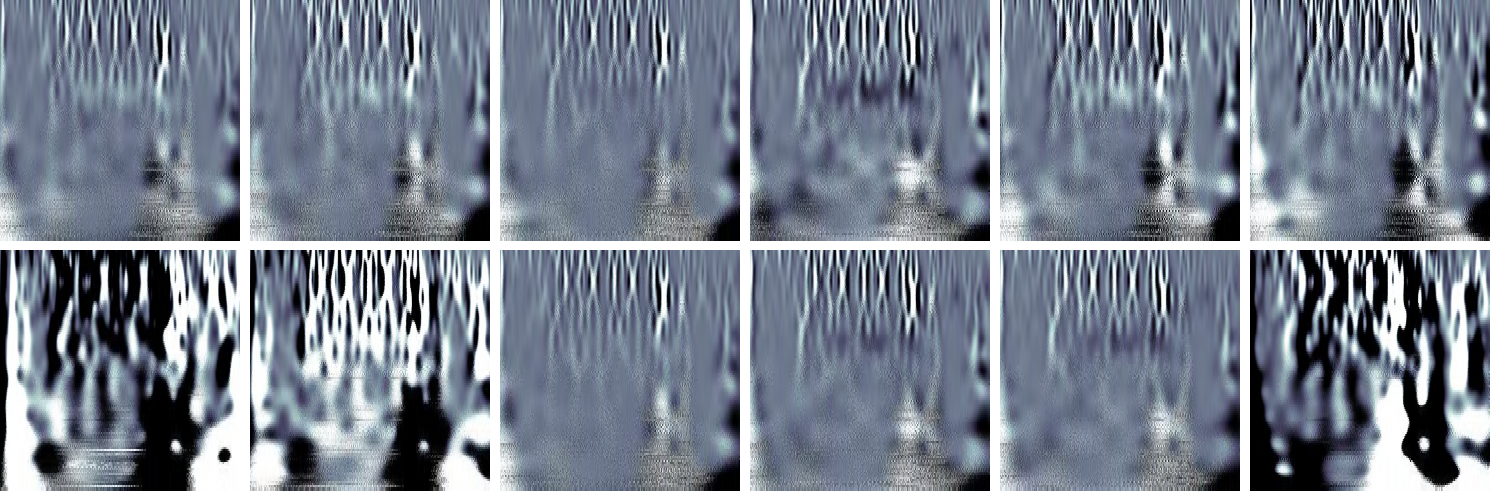}
\caption{ \label{fig:packet_comp}  Difference scalogram, $\Delta_{ab}$, for signal $m=1$ as function of transmitter index (1--12). Transmitters from left to right are Y06v2 Tx1,Y06v2 Tx2,R05v1 Tx1, R05v1 Tx2,R04v1 Tx1,R04v1 Tx2 for the top row, and R03v1 Tx1, R03v1 Tx2,Y04v2 Tx1,Y04v2 Tx2,Y10v2 Tx1,Y10v2 Tx2 for the bottom row.    It can be seen from these scalograms that different features can be identified for each transmitter.   }
\end{figure}

\begin{figure}[h]
\includegraphics[width=\columnwidth]{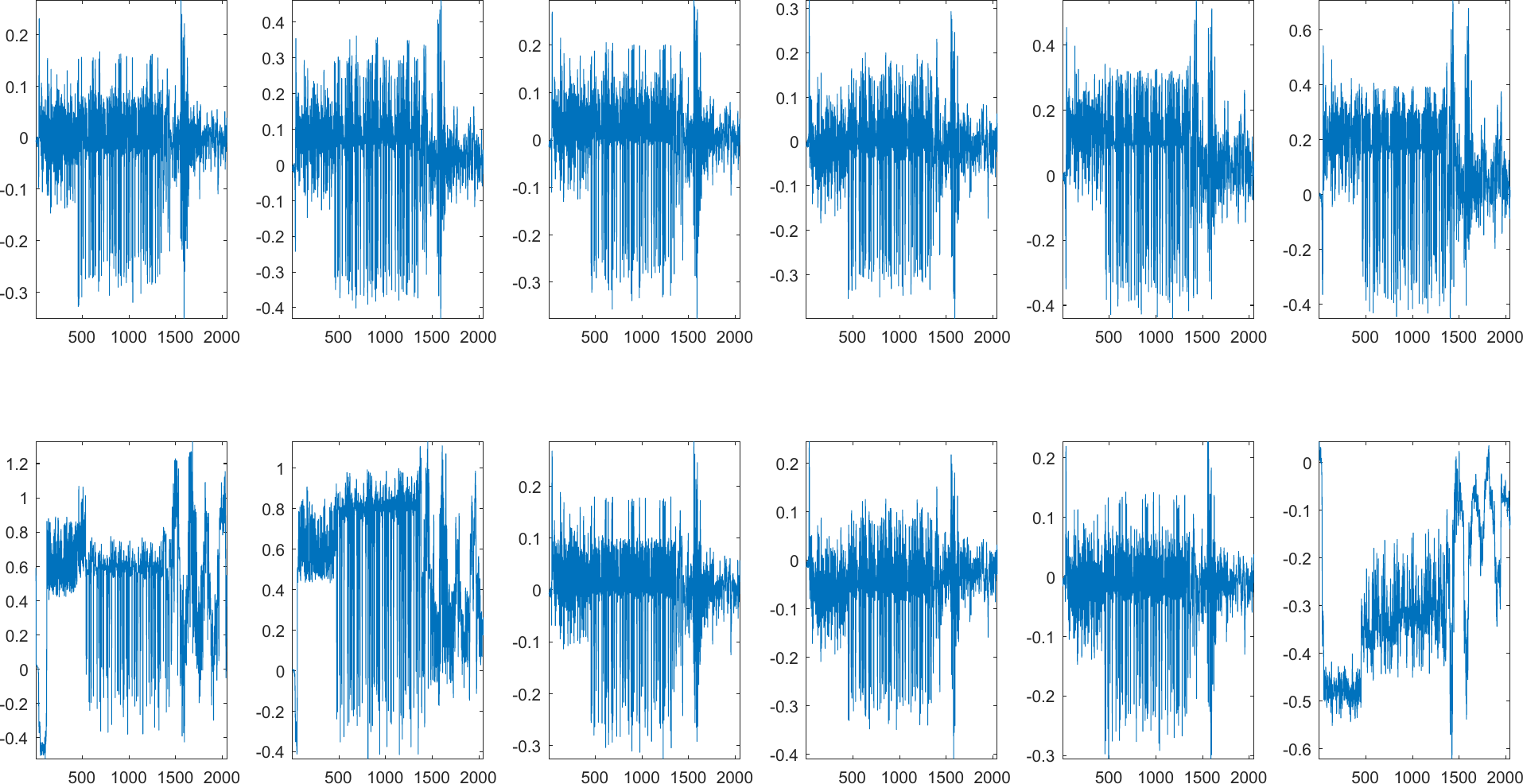}
\caption{ \label{fig:TD_comp} Time-domain of the initial part of the difference RF signal $\delta_i$ for signal 1 broadcast across 12 different transmitters.   In this time-domain representation, it is much more difficult to identify features that are unique to each transmitter (c.f., Figure~\ref{fig:packet_comp}).     }
\end{figure}

\begin{figure*}
\includegraphics[width=0.3\textwidth]{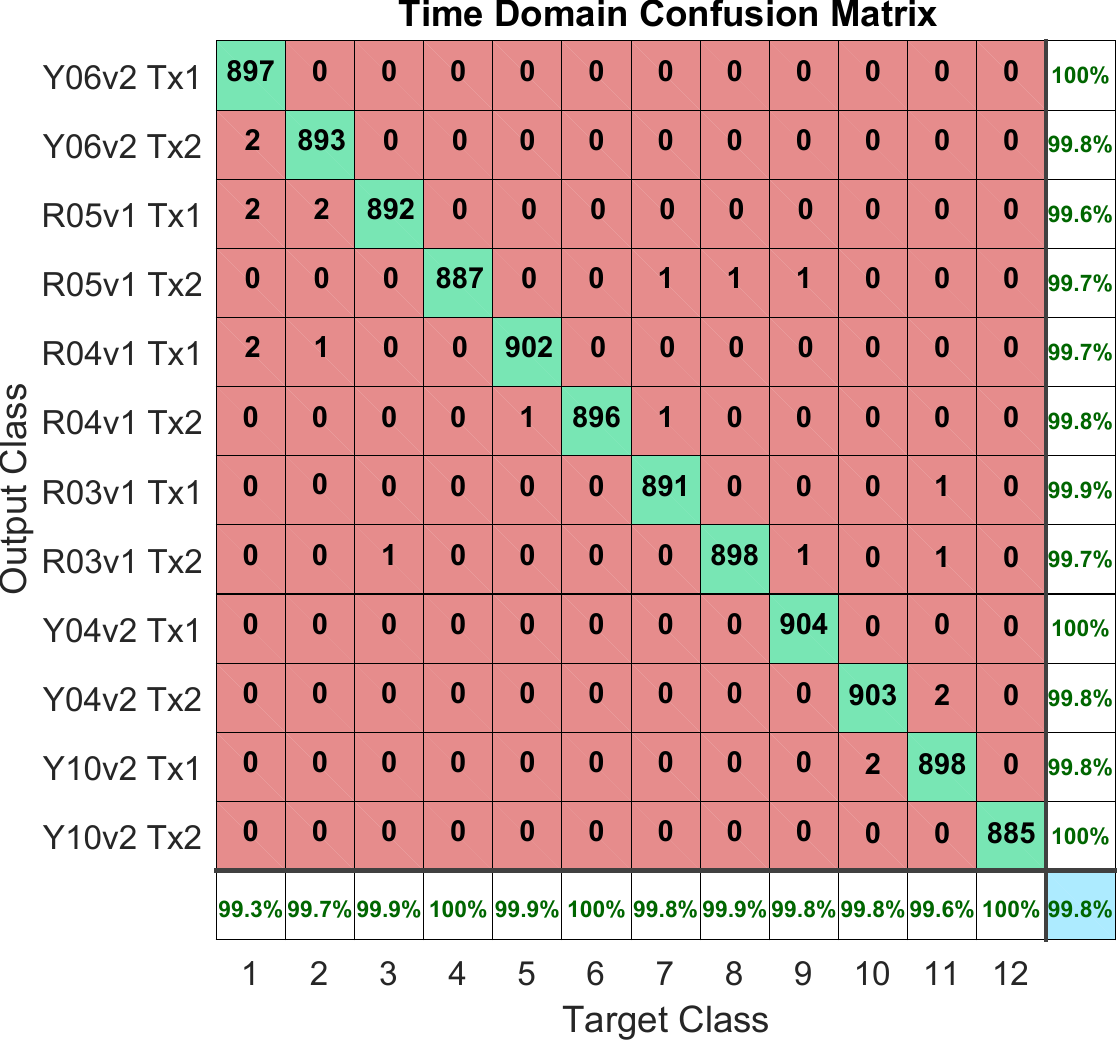}
\includegraphics[width=0.3\textwidth]{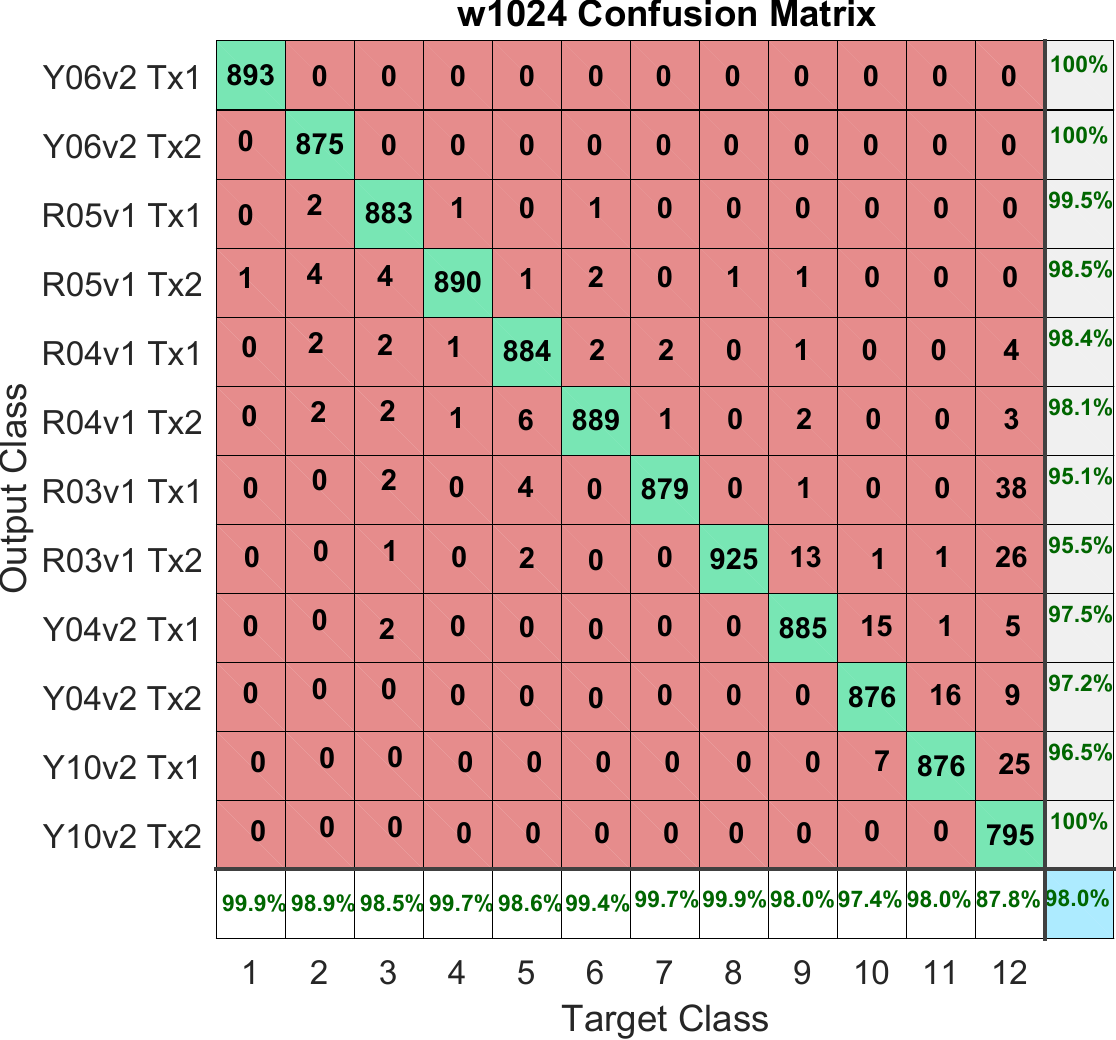}
\includegraphics[width=0.3\textwidth]{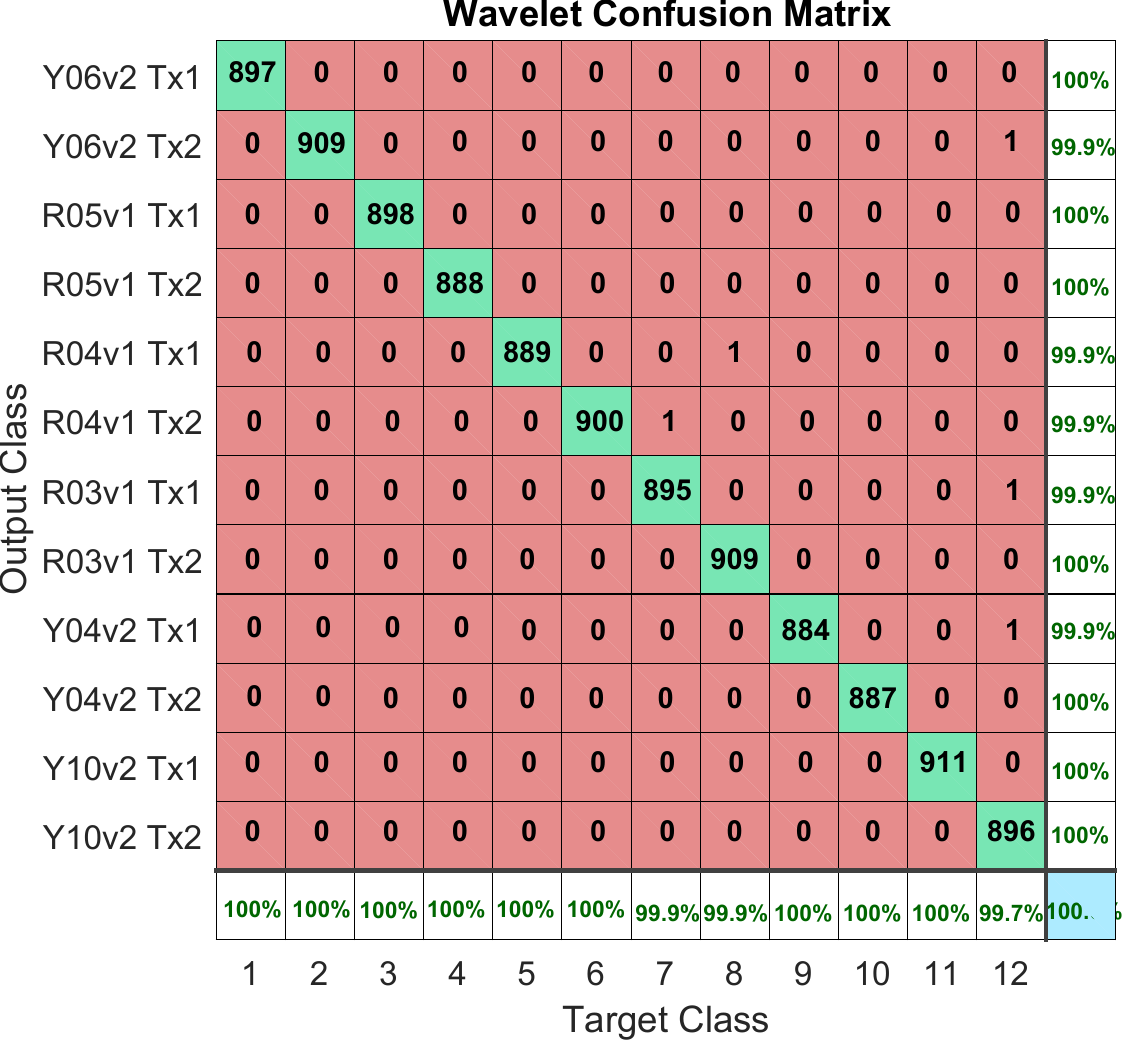}
\caption{ \label{fig:conf_comp}   Confusion matrices in the case of wavelet preconditioning (right) vs time-domain methods (left, middle).   The wavelet preconditioning method outperforms both time-domain methods by achieving nearly 100\% accuracy with 12 transmitters.  Confusion matrices plotted using the MATLAB command \texttt{plotconfusion}~\cite{matlab:plotconfusion}; the rows and columns do not always add up exactly to 100\% because the data is randomly selected from a pool of all signals, so some transmitters may get a little less or more signals for different runs.
  }
\end{figure*}

The CWT of $\vec{v}$ was computed using the Morlet wavelet to yield a scalogram stored as a 128$\times$2,048 matrix (128 scales, 2,048 translations).   The scalogram is denoted as $s_{abtm}$, where $a,b$ indices denote the scale and translation index, $t\in\{1,2,\dots,12\}$ is the transmitter index and $m\in\{1,2,\dots,30\}$ is the signal number (out of the 1,000 signals we collected, only 30 were used to compute the variances to save time).  The variance of the $a,b$-th pixel in the scalogram was computed as 
 $$var(s_{ab})=\frac{1}{N_t N_s-1} \sum_{t=1}^{N_t} \sum_{m=1}^{N_s}(s_{abtm}-\overline{s_{ab}})^2,$$ 
where $N_t=12$ is the number of transmitters, $N_s=30$ is the number of signals ($N_t N_s=360$ packets in total) and 
$$\overline{s_{ab}}=\frac{1}{N_t N_s} \sum_{t=1}^{N_t} \sum_{m=1}^{N_s} s_{abtm}.$$  
The four variance maps are shown in Figure~\ref{fig:var_comp}.   As can be seen, the variance of the absolute value plot shows excellent sparsity compared to the other plots, suggesting a potentially useful representation of the RF signal for feature extraction via a handful of small regions of the map.   On the other hand, variance is spread more uniformly across the map for the remaining cases. 

{\bf The question we want to address next is whether or not the CWT scalogram (2D) is better at transmitter identification than the time-domain signal (1D).} The superiority of the wavelet transform representation for this RF application can be  illustrated by comparing the scalogram vs time-domain signal for an identical  signal that is sent across different transmitters.  In  Figure~\ref{fig:packet_comp}, we show scalograms for signal 1 sent across 12  different transmitters.  To highlight only the ``fluctuations'' relative to  some convenient mean, we plot the difference  $\Delta_{ab}=s_{abtm}-\overline{s_{ab}}$ at each pixel $a,b$, where  $t=1,\dots,12$ and $m=1$ (signal 1).

The analogous comparison for the time-domain signal is shown in Figure~\ref{fig:TD_comp}, where the time-domains of signal 1 broadcast across the 12 different transmitters are compared.    If $f_{itm}$ denotes the magnitude of the complex-valued time-domain RF signal, $i=1,\dots,2048$ is the time index, $t=1,\dots,N_t$ ($N_t=12$) is the transmitter index and $m=1,\dots,N_s$ ($N_s=30$) is the signal number, the mean across signals and transmitters is 
$$ \overline{f_i} = \frac{1}{N_t N_s} \sum_{t=1}^{N_t} \sum_{m=1}^{N_s} f_{itm}. $$ 
In Figure~\ref{fig:TD_comp}, it is the difference signal $\delta_i=f_i-\overline{f_i}$ that is plotted to highlight the shifts relative to baseline.  While certain differences can be seen among certain transmitters, the wavelet representation (Fig.~\ref{fig:packet_comp}) does a better job at providing a transmitter fingerprint thanks to the spatial correlations introduced by the transform.  Such correlations in the signal are much more difficult observe in the time-domain (Fig.~\ref{fig:TD_comp}) representation.

Next, we evaluated the performance of the wavelet feature learning method by feeding the wavelet-preconditioned signals into the MST system and comparing against the time-domain methods.  Figure~\ref{fig:conf_comp} shows confusion matrix for results obtained under three different scenarios.  In the first scenario, training is done with time domain signal using method 1, where real and imaginary components from an initial segment of the RF signal following the onset (w1024) are concatenated and sent to MST. This scenario leads to the ``w1024 confusion matrix'' (middle).

In the second scenario, training is done with time-domain signal similar to method 1 but this time, the absolute is used instead of concatenating real and imaginary parts. This is labeled as ``time domain confusion matrix'' (left). In the third scenario, MST training is done with features extracted from the wavelet transform as described previously.  The wavelet preconditioning (with SOM and pooling layers, as shown in Fig.~\ref{fig:front_end}) is fed to the MST.  The results are shown in the ``wavelet confusion matrix'' (right).  As can be seen, the wavelet preconditioning outperforms both time-domain methods by achieving nearly 100\% accuracy in the case of 12 transmitters. Note that for these results, only 1,024 samples after the onset (instead of 2,048) were used by CWT for fair comparison.

\begin{table}[h]
\centering

\begin{tabular}{lrr}
  \hdcell{Method}&\hdcell{Accuracy (avg)}&\hdcell{Time (rel.)} \\
  Time-Domain w1024 & 52.1\% & 1.7 \\
  Wavelet Precond.  & 93.3\% & 1.0 \\
\end{tabular}

\caption{Wavelet preconditioning vs time-domain method for w1024.  The wavelet method outperforms the time-domain method in terms of accuracy and speed. (Note: this is for a single MLP, not MST; Table~\ref{tab:table_L1a} results are for MST.) Training times are given in arbitrary (relative) units.    \label{tab:single_MLP} } 
\end{table}

When scaling up to large numbers of transmitters, it will be critical to use an appropriate feature extraction method that will distill the essential features that are unique to each transmitter.  Here we show that the addition of wavelet preconditioning leads to higher accuracy compared to the time-domain method. To compare wavelet and time-domain methods, we calculated the average result of 10 runs with a single MLP with 2 hidden layers/100 neurons per layer and first order training. The results presented in Table~\ref{tab:single_MLP} show both the average performance and the average convergence time.  

\subsection{Incremental Learning with Wavelets \label{sec:scale-wprec}}

In this section we ask {\bf whether wavelet preconditioning can easily extend to capture new devices.}

We repeated the experiment from the incremental learning section (\secref{incremental}) for a larger number of transmitters, using wavelet preconditioning as the data preparation method. Here, only data from $k$ out of $n$ transmitters was used to train the first stage. Data from the remaining transmitters was then introduced in the second and third stages. The results for several training and testing partitioning percentages, and different $n$/$k$ ratios are shown in Table~\ref{tab:table_L2b}. We note that even under extremely severe conditions (1\% training and $n$/$k$ = $12$/$3$) the system still maintains a remarkably high performance.  Thus, we conclude that wavelet preconditioning of MST is the most promising approach for transmitter identification and classification investigated to date.

\begin{table}[h]
\noindent
\begin{tabular}{ccc}
\hdcell{Training/Testing} & \hdcell{$n$/$k$} & \hdcell{\bf Accuracy} \\
  90/10 & 12/6 & 100\% \\
 50/50 & 12/6 &100\% \\
  10/90 & 12/6 & 99.95\% \\
1/99 & 12/3 & 94.45\% \\
\end{tabular}
\caption{ \label{tab:table_L2b} Incremental learning results with datasets obtained using wavelets preconditionning. }
\end{table}

\section{Conclusion}

Our results show that a new ANN strategy based on second-order training of MST is well-suited for RF transmitter identification and classification problems.  MST outperforms the state-of-the-art ML algorithms DLL, CNN and SVM in RF applications.  We also found that wavelet preconditioning enabled us to not only get higher  (up to 100\%) accuracy but reduce the complexity of identifying a large number of unknown transmitters.  We anticipate that this scalability property will enable ML identification of a very large number of unknown transmitters and assign a unique identifier to each.  We note in closing that while the results are promising, this study should be viewed as a proof-of-concept study until it is extended to the more challenging conditions encountered in real busy environments.  The next obvious steps would involve increasing the number of transmitters, testing the robustness of the method with varying packets, noisy channels, under conditions of overlapping transmissions, interfering channels, moving sources (Doppler effect), jamming and other channel effects
added.



\bibliographystyle{unsrt} 
\bibliography{references}

\end{document}